\documentclass{jfm}
\usepackage{graphicx,ulem}
\usepackage{epstopdf, epsfig}
\usepackage{amsmath}
\usepackage{amsfonts}
\usepackage{amssymb}

\usepackage{xcolor}

\usepackage{mathrsfs}

\newcommand{\diff}{\text{d}}

\usepackage{pgfplots}
\usepgfplotslibrary{groupplots}
\usepgfplotslibrary{fillbetween}
\pgfplotsset{compat=newest}
     \newlength\fheight 
     \newlength\fwidth
     \newlength\svgwidth

\shorttitle{Convection in a mushy-layer along a heated wall}
\shortauthor{S. Boury, C.R. Meyer, G.M. Vasil and A.J. Wells}

\title{Convection in a mushy-layer along a heated wall}

\author{S. Boury\aff{1}
  \corresp{\email{sb7918@nyu.edu}}, 
  C.R. Meyer\aff{2},
  G.M. Vasil\aff{3},
  \and A.J. Wells\aff{4}}

\affiliation{\aff{1}Courant Institute of Mathematical Sciences, New York University, New York, NY 10012, USA
\aff{2}Thayer School of Engineering, Dartmouth College, 14 Engineering Drive, Hanover, NH, 03755, USA
\aff{3}School of Mathematics and Statistics, University of Sydney, Sydney,
New South Wales 2006, Australia
\aff{4}Atmospheric, Oceanic and Planetary Physics, Department of Physics, Clarendon Laboratory, University of Oxford, Parks Road, Oxford OX1 3PU, UK}

\begin{document}

\maketitle

\begin{abstract}
	
	Motivated by the mushy zones of sea ice, volcanoes, and icy moons of the outer solar system, we perform a theoretical and numerical study of boundary-layer convection along a vertical heated wall in a bounded ideal mushy region. The mush is comprised of a porous and reactive binary alloy with a mixture of saline liquid in a solid matrix, and is studied in the near-eutectic approximation. Here we demonstrate the existence of four regions and study their behavior asymptotically. Starting from the bottom of the wall, the four regions are (i) an isotropic corner region; (ii) a buoyancy dominated vertical boundary layer; (iii) an isotropic connection region; and (iv) a horizontal boundary layer at the top boundary with strong gradients of pressure and buoyancy. Scalings from numerical simulations are consistent with the theoretical predictions. Close to the heated wall, the convection in the mushy layer is similar to a rising buoyant plume abruptly stopped at the top, leading to increased pressure and temperature in the upper region, whose impact is discussed as an efficient melting mechanism.
	
\end{abstract}

\begin{keywords}
Authors should not enter keywords on the manuscript, as these must be chosen by the author during the online submission process and will then be added during the typesetting process (see http://journals.cambridge.org/data/\linebreak[3]relatedlink/jfm-\linebreak[3]keywords.pdf for the full list)
\end{keywords}

% ################################################################################ %
\section{Introduction}
		
		Convection of pore fluid in partially molten porous media arises in numerous environmental and industrial systems \citep{worster1997}, such as brine drainage in sea ice \citep{hunke2011}, solidification of magma \citep{tait1992}, and freckle formation in alloys \citep{fowler1985}. An important class of problems arise when solidifying alloys are cooled from boundaries \citep{worster1986, worster2000}. Mushy zones cooled from the boundaries are found in many geological phenomenon, such as hot volcanic dikes \citep{furumoto1975, cheng1977} and chimneys in analogues of metalic alloy solidification \citep{copley1970}. When imposed to a partially solidified region, boundary cooling can trigger free convection of the liquid flowing through the porous solid mushy zone. This has been studied in the case of solidification with cooling from a horizontal boundary \citep[e.g.][]{worster1997, worster2000, anderson2019}, but also in the more complex case of cooling at a vertical boundary \citep{huppert1990, guba2006}. Similar types of convective boundary layers from vertically-distributed sources have been studied in non-reactive porous media \citep{cheng1977}, and for convection adjacent to boundaries undergoing phase change \citep[e.g.][]{carey1982, bloomfield2003}. Free convection is also important when the flow is internal to the mush and coupled to the solidification process \citep[e.g.][]{guba2006}. We here focus on convective boundary layer flows next to a heated vertical boundary, or vertical planar buoyancy source in a reactive porous mushy layer.

		One particular application of the mushy zone surrounding a heated wall is as a model for shear heating induced melting along tectonic features in the shells of icy satellites, such as the tiger stripes of Enceladus \citep{nimmo2007} or Europa \citep{hammond2019}. Prevailing models of heating at localised fractures in ice shells have treated pure fresh ice \citep{gaidos2000, han2008}, but the ice shell may be a binary mixture of salts and water \citep[e.g.][]{mccord1999}. Partial melting of the salt and ice mixture can result in the formation of a mushy layer, which is a reactive porous layer of fresh ice crystals and saline liquid brine, and allows for convection and heat transport in the interstitial fluid \citep{worster1997, wells2019}. Tidally induced shear heating along the fracture could cause the region around the fracture to warm above the eutectic temperature \citep{nimmo2007} and partially melt to form a mushy zone with two-phase coexistence of salty liquid in a solid ice matrix. The liberated brine (liquid water and dissolved salt) will convect due to variations in salt content through the mushy zone and the motion of the mobile brine could provide the source for the observed Enceladus south polar plume.
		
		Convective flows in porous regions regions sustained by heat provided by a vertical wall have been shown to be described by self-similar scalings. For example, \cite{cheng1977} and \cite{ingham1986} considered a heated vertical plate in a semi-infinite domain in a pure saturated porous media. Depending on the thermal input at the wall, the flow and temperature in the porous medium can be described by a self-similar solution within a thermal boundary layer \citep{ingham1986}.\cite{guba2006} considered convection in a mushy zone at an isothermal vertical boundary at the eutectic freezing front formed during horizontal directional solidification. They showed that the flow close to the wall in a semi-infinite mushy region can also be described by a self-similar solution. By contrast, we here consider the buoyant flow generated by melting of a mushy layer from a vertical boundary with an imposed constant heat flux, in particular investigating the flow dynamics in a domain of finite height.
		
		In this paper, we report a theoretical and numerical study of free convection along a heated wall that imposes a heat flux into a partially melted mushy region of finite depth. We focus on a reduced model using the near-eutectic approximation \citep{fowler1985, worster1997} for a mushy layer that is already sufficiently warm to allow partial melt. Our model yields dynamical insight into flow patterns in an existing region of porous mush, but does not account for the initial transient production of the melted region. We study the flow in a finite depth closed box and focus primarily on the flow close to the heated wall, ignoring possible far-field feedbacks such as flow-boundary interactions and varying buoyancy due to the return flow. In section~\ref{sec:model} we describe the model and present the governing equations as well as the numerical methods. We identify four different regimes near the wall (from bottom to top: an isotropic diffusive tip; a vertical buoyant boundary layer; an isotropic transition region; and a connection region with scaling inherited from a horizontal boundary layer). These different regimes predicted by the model are detailed in section~\ref{sec:results} in which the theoretical derivation is supported by a quantitative check of the proposed scalings in numerical simulations of flow in a porous media consistent with the near-eutectic approximation to the mushy layer dynamics. Our conclusions and discussion are presented in section~\ref{sec:conclusion}.

% ################################################################################ %

\section{Model}
\label{sec:model}
	
% -------------------------------------------------------------------------------- %
	\subsection{Governing Equations}
	
		\begin{figure}
			\centering
			\includegraphics[scale=1]{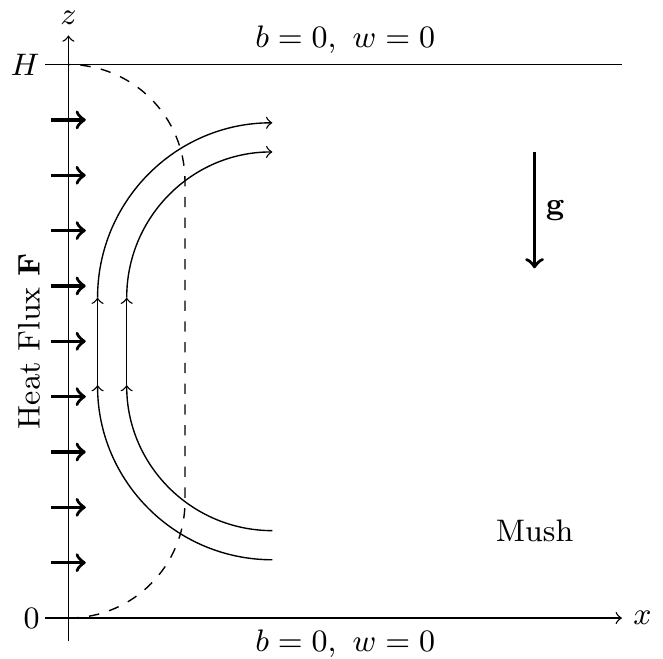}
			\caption{Formulation of the problem: a full depth wall, from $z=0$ to $z=H$, is heated, imposing a constant heat flux $\mathbf{F}=F\mathbf{e_x}$ to the mushy region. A thermal boundary layer develops close to the wall, in which the convective flow is studied. The buoyancy $b$ and vertical velocity $w$ are fixed at zero at the top and bottom.}
			\label{fig:modconvection}
		\end{figure}
	
		We consider a binary alloy that constitutes an ideal mushy region in a two-dimensional domain (see figure~\ref{fig:modconvection}), i.e. a region in which the solid-liquid interface has become so convoluted that the solid forms a porous matrix of fresh ice crystals saturated by brine \citep{worster2000}. Here we assume that the material properties are the same in each phase. One such example of a mushy layer is porous sea ice (see \cite{wells2019} and references therein).  A vertical boundary is located at $x=0$ and we impose a constant horizontal heat flux $\mathbf{F}=F\mathbf{e_x}$ into the mush. In the mushy zone, the governing equations for the temperature $T$, the liquid salinity $S$, the solid fraction $\phi$, and the velocity field $\mathbf{u} = (u,0,w)$ are given by~\citep{worster1997}
			\begin{eqnarray}
				\dfrac{\partial T}{\partial t} + (\mathbf{u}\cdot\mathbf{\nabla}) T &=& \kappa \mathbf{\nabla}^2 T + \frac{L}{C_p} \dfrac{\partial \phi}{\partial t},\label{eq:set1} \\
				(1-\phi)\dfrac{\partial S}{\partial t} + (\mathbf{u}\cdot\mathbf{\nabla}) S &=& S \dfrac{\partial \phi}{\partial t},\label{eq:set2} \\
				T &=& T_0 - m (S-S_0),\label{eq:set3} \\
				\frac{\mu}{\Pi} \mathbf{u} &=& -\mathbf{\nabla} P + \rho \mathbf{g}, \label{eq:set4}\\
				\mathbf{\nabla}\cdot \mathbf{u} &=& 0.\label{eq:set5}
			\end{eqnarray}
			Equations~\eqref{eq:set1} and~\eqref{eq:set2} are the heat and salinity advection-diffusion equations taking into account the phase change correction, where $L$ is the latent heat, $C_p$ the heat capacity, and $\kappa$ the thermal diffusivity. Due to the relatively low Schmidt number (ratio of solute and thermal diffusivity) in the considered systems, we neglect the diffusive term in equation~\eqref{eq:set2}. The liquidus equation~\eqref{eq:set3} is a closure relation for the system when considering an ideal mush in which the liquid and solid phases are at equilibrium and aligns on the liquidus of the binary alloy. In the mush, we assume that the temperature and solute concentration or salinity are linked by a linear relation with $S_0$ a reference solute concentration, $T_0$ the freezing temperature at $S_0$, and $m$ a constant coefficient~\citep{worster1986, worster2000}. The brine flow through the porous mush can be modeled using Darcy's law~\eqref{eq:set4}, where $\mu$ is the viscosity, $\Pi$ the permeability of the porous media, $\rho$ the fluid density, $\mathbf{g}$ the gravitational acceleration, and $P$ the pressure field. For the purpose of this study, we assume that the permeability $\Pi$ of the mush is constant and that there is no feedback from the temperature and velocity field on it. Equation~\eqref{eq:set5} is the continuity equation for mass conservation in the flow when assuming incompressibility.
	
		In Darcy's law \eqref{eq:set4}, the density of the fluid $\rho$ is a function of the temperature and of the salinity, which we approximate as
		\begin{equation}
			\rho(T,S) = \rho_0 \left[ 1 - \alpha_T (T-T_0) + \alpha_S (S-S_0) \right],
		\end{equation}
		where $\rho_0$ is a reference density, $\alpha_T$ and $-\alpha_S$ are the coefficients of thermal and solutal expansion, respectively, and $T_0$ and $S_0$ are reference temperature and salinity. From the liquidus equation \eqref{eq:set3}, however, temperature and salinity are linearly related through a constant coefficient $m$, and the density $\rho$ can be written as a function that depends only on temperature. We have
		\begin{equation}
			\rho(T) = \rho_0 \left[ 1 - \alpha (T-T_0) \right],
		\end{equation}
		with $\alpha = \alpha_T + \alpha_S / m$. The coefficient $\alpha_S / m$ often dominates, in which case the buoyancy is controlled by the meltwater released by phase change as the system rapidly relaxes to local thermal equilbrium.
		
		We now define the buoyancy $b$ as a linear function of temperature
		\begin{equation}
			b = \alpha g \rho_0 (T-T_0),
		\end{equation}
		so that
		\begin{equation}
			\rho g = \rho_0 g - b.
		\end{equation}
		where $g=|\mathbf{g}|$. Using these definitions, we write both the advection-diffusion equation for heat \eqref{eq:set1} and the advection-diffusion equation for salinity \eqref{eq:set2} in terms of $b$. Note that in Darcy's law, the constant term $\rho_0 \mathbf{g} $ can be included as the gradient of a hydrostatic pressure field.
		
		The buoyancy can be fully determined using boundary conditions that close the system. We here consider an imposed heat flux condition at $x=0$ as
		\begin{equation}
			F = -\lambda \dfrac{\partial T}{\partial x} = - \frac{\lambda}{\alpha g \rho_0} \dfrac{\partial b}{\partial x},\label{eq:BC2.10}
		\end{equation}
		where $F$ is a constant heat flux, parametrising the problem, and $\lambda$ the thermal conductivity of water, assumed to be the same in the liquid and the ice. Note that by symmetry this yields the same thermal forcing as a localised delta-function line source of heating at the centre of a domain of twice the width, with magnitude $2F$. This will be later modeled using a gaussian approximation to the delta function in the numerical solutions. In the far-field, the buoyancy and the velocities are assumed to go to zero, and we also assume zero vertical velocity and zero buoyancy at the top and bottom boundaries. Although this model neglects some physical processes relevant to the dynamics of geophysical settings, such as temporally and spatially intermittent heating sources, the goal is to yield an analytically tractable problem capturing the key elements of the convective boundary layer flow near to a heated vertical boundary in order to build initial insight.

% -------------------------------------------------------------------------------- %
	\subsection{Dimensionless Equations}
	
		We consider a characteristic length scale $H$, corresponding to the vertical size of the wall, and a characteristic time scale $H^2/\kappa$, where $\kappa$ is the thermal diffusivity. We therefore construct a characteristic velocity scale $\kappa / H$. This leads us to the definition of dimensionless lengths, time, and velocities
		\begin{equation}
			x ~\rightarrow~ \tilde{x} = \frac{x}{H}, \mathrm{~~~~~~~} z ~\rightarrow~ \tilde{z}=\frac{z}{H},\mathrm{~~~~~~~and~~~~~~~}t ~\rightarrow~ \tilde{t} = \frac{\kappa t}{H^2},
		\end{equation}
		and
		\begin{equation}
			u ~\rightarrow~ \tilde{u} =\frac{H u}{\kappa} \mathrm{~~~~~~~and~~~~~~~} w~\rightarrow~ \tilde{w} =\frac{H w}{\kappa}.
		\end{equation}
		A similar scaling can also be found for the buoyancy and the pressure
		\begin{equation}
			b ~\rightarrow~ \tilde{b} = \frac{H \Pi_0 b}{\kappa \mu}\mathrm{~~~~~~~and~~~~~~~}P ~\rightarrow~ \tilde{p} = \frac{\Pi_0 P}{\kappa \mu} + \frac{g H \Pi_0 \rho_0}{\kappa \mu} \tilde{z}.
		\end{equation}
		For the sake of brevity, the dimensionless variables will be noted without tildes. Hence, the dimensionless equations are
		\begin{eqnarray}
				\dfrac{\partial b}{\partial t} + (\mathbf{u}\cdot\mathbf{\nabla}) b &=& \mathbf{\nabla}^2 b + \mathscr{S} Ra \dfrac{\partial \phi}{\partial t}, \label{eq2:14}\\
				\dfrac{\partial}{\partial t}\left[\mathscr{C} Ra \phi + b (1-\phi) \right] + (\mathbf{u}\cdot\mathbf{\nabla}) b &=& 0,\label{eq2:15}\\
			\mathbf{u} &=& -\mathbf{\nabla} p + b\hat{\bf{z}}, \label{eq2:16}\\
			\mathbf{\nabla}\cdot \mathbf{u} &=& 0. \label{eq2:17}
		\end{eqnarray}
		where $\mathscr{S}$ is the Stefan number defined as
		\begin{equation}
			\mathscr{S} = \frac{L \lambda}{C_p F H},\label{eq:stephan}
		\end{equation}
		$\mathscr{C}$ is the compositional ratio
		\begin{equation}
			\mathscr{C} = \frac{m S_0 \lambda}{F H},\label{eq:comporation}
		\end{equation}
		and the porous medium Rayleigh number Ra is
		\begin{equation}
			Ra = \frac{\alpha g \rho_0 \Pi_0 F H^2}{\kappa^2 \mu C_p},
		\end{equation}
		that represents the ratio of buoyant to dissipative mechanisms (thermal and viscous).
		
		This form of Rayleigh number emerges from the dimensionless form of the heat flux boundary condition~\eqref{eq:BC2.10} expressed in terms of the buoyancy $b$
		\begin{equation}
			F = - \lambda \dfrac{\partial T}{\partial x} ~\rightarrow~ \frac{\alpha g \rho_0 \Pi_0 H^2}{\kappa^2 \mu C_p} F = -\dfrac{\partial \tilde{b}}{\partial \tilde{x}},
		\end{equation}
		which yields
		\begin{equation}
			-\dfrac{\partial b}{\partial x} = Ra \mathrm{~~~~~at~~~~} x=0, \label{eq:RaBC}
		\end{equation}
		with dimensionless variables and tildes dropped. In the numerical simulations, we therefore discuss the heat flux $F$ and the Rayleigh number $Ra$ interchangeably. It can also be used to defined a natural length scale $h^\star$ over which the buoyancy and dissipative mechanisms have commensurate magnitude written as
		\begin{equation}
			h^\star = \sqrt{\frac{H^2}{Ra}} = \left(\frac{\kappa^2 \mu C_p}{\alpha g \rho_0 \Pi_0 F}\right)^{1/2}.
			\label{eq:h}
		\end{equation}
		We then define the dimensionless length scale $h=h^\star/H$	.

% -------------------------------------------------------------------------------- %
	\subsection{Near-Eutectic Approximation}
		
		Similarly to the analysis of a freezing front in \cite{guba2006}, we now consider an approximation that is similar to the near-eutectic approximation described by~\cite{fowler1985} and~\cite{worster1986}. The relevant limit considers $\mathscr{C} \gg 1$, which corresponds to a relatively weak magnitude of heating versus the freezing point depression for the background composition. This is similar to the near-eutectic approximation, where composition or temperature differences across the system are assumed to be a small fraction of the eutectic composition or temperature difference between the eutectic point and freezing  point of pure liquid. If the largest lengthscales in the system are of order 1, then condition~\eqref{eq:RaBC} implies an order of magnitude upper bound of $b=\mathcal{O}(Ra)$. We then consider a limit where $\mathscr{C}\gg 1$, and $\phi = O(1/\mathscr{C}) \ll 1$. Approximating $(1 - \phi) b \simeq b$ at leading order in~\eqref{eq2:15} yields to an equation for $\partial_t \phi$ in terms of $\partial _t b$. Then, eliminating $\partial_t \phi$ from~\eqref{eq2:14} results in~\eqref{eq:advecdiffb}, an advection-diffusion equation for the buoyancy that does not involve the solid fraction $\phi$
		\begin{equation}
			\Omega \left[ \dfrac{\partial b}{\partial t} + (\mathbf{u}\cdot\mathbf{\nabla}) b \right] = \mathbf{\nabla}^2 b,
			\label{eq:advecdiffb}
		\end{equation}
		where $\Omega = 1 + \mathscr{S} / \mathscr{C}$ represents a modified dimensionless heat capacity, which accounts for the impact of phase changes and  latent heat transfer that occurs during warming or cooling of the solid matrix \citep{huppert2012}. Note that in this approximation, $\phi$ evolves as a slaved variable, with no impact on $b$ or $\mathbf{u}$.  The reduced system~\eqref{eq2:15}, \eqref{eq2:16}, and~\eqref{eq:advecdiffb} hence characterises buoyancy driven convective flow in a porous medium with modified heat capacity.

% -------------------------------------------------------------------------------- %
	\subsection{Numerical Methods}
	
\begin{figure}
			\centering
			\includegraphics[scale=0.9]{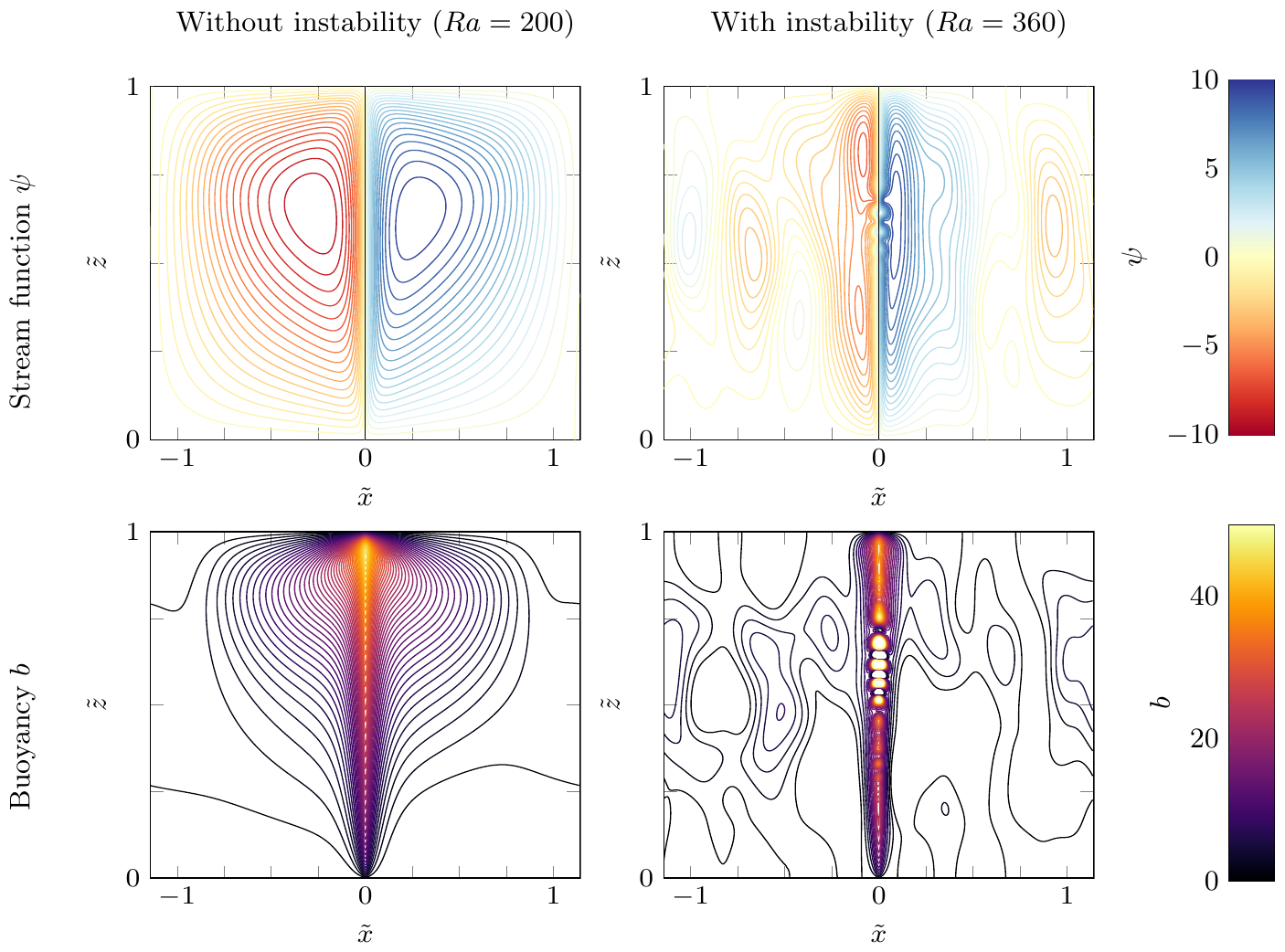}
			\caption{Contours of stream function (top row) and buoyancy (bottom row) in a DNS using Dedalus \citep{burns2019}, imposing a $z$-independent localised heat source around $\tilde{x}=0$ which emulates a heated wall, and zero buoyancy and vertical veloctity at the top and at the bottom of the domain. Contour values go from $-10$ to $+10$ with a spacing of $0.5$. Blue is clockwise and red is anti-clockwise. Left: Steady state reached at relatively low heating $Ra=200$ and $\Omega=1$. Right: Instability triggered with increased heating, at $Ra=360$ and $\Omega=1$. We do not consider these larger and unstable Rayleigh number solutions further here.}
			\label{fig:psi}
		\end{figure}

	We run Direct Numerical Simulations (DNS) of the governing equations \eqref{eq2:16}, \eqref{eq2:17} and \eqref{eq:advecdiffb} using a pseudospectral method implemented in Dedalus \citep{burns2019}. We use a two-dimensional horizontally periodic domain $(x,z)\in [-L_x /2,~+L_x/2] \times [0,~H]$, with $256\times 256$ nodes using a Fourier basis in the $x$ direction and a Chebyshev basis in the $z$ direction which naturally clusters grid points near the boundaries. The resolution and the time step have been set to ensure the convergence of the simulation, defined by the saturation of the total kinetic energy of the system with an uncertainty of $10^{-4}$. Boundary conditions of no orthogonal flow and zero buoyancy are applied at the top and bottom of the domain. Instead of applying the flux boundary condition~\eqref{eq:RaBC} directly, at $x=0$, we use an approximation to a line source, with a horizontal Gaussian heating pattern that is constant along the vertical wall,
	\begin{equation}
		Q(x,z) = \frac{Q_0}{\sqrt{2 \pi \sigma^2}} \exp \left( - \frac{x^2}{2 \sigma^2}\right),
	\end{equation}
	where $Q_0 = 2F$ is the heating coefficient and $\sigma$ the width of the Gaussian. We solve equations~\eqref{eq2:16} through~\eqref{eq:advecdiffb}, which corresponds to the dimensionless set of equations for porous media convection, integrating out to the steady state reached when convergence of the code is observed, i.e. when the total kinetic energy reaches a plateau. Example results are presented in figure~\ref{fig:psi} to provide context for the subsequent discussion.
	
	The left pannel in figure~\ref{fig:psi} shows contours of the stream function in the steady state, in a DNS with $L_x = 16$ and $H = 7$ (size of the wall, in nondimensional length units), $Ra = 200$, and $\sigma/H=7.1 \times 10^{-3}$ (width of the Gaussian, in length units). Two convective cells are created, one flowing clockwise (in blue) and one flowing anti-clockwise (in red). They are symmetric right-left but the up-down symmetry is broken because of buoyancy and gravity. From the bunching of streamlines in the thin layer close to the heated wall, we can infer the existence of a vertical boundary layer in which the flow is accelerated vertically to the top. The right pannel in figure~\ref{fig:psi} shows the instability triggered when the heating is increased further, to $Ra = 360$. Here we focus on the steady solutions, hence the study of this instability is beyond the scope of this work.
	
% ################################################################################ %
\section{Results}
\label{sec:results}

	We focus primarily on the boundary layer flow near the wall, supplied by a reservoir of far-field mush. The specific pattern of the far-field return flow may depend more strongly on the far-field boundary conditions - in geophysical applications this might include the geometry of any mush-eutectic phase boundary. But one would expect the leading order characteristics of the near-wall boundary-layer flow to depend primarily on conditions at the heated wall and immediately outside the boundary layer, and be less sensitive to details of the far-field return flow.

\begin{figure}
			\centering
			\includegraphics[scale=1]{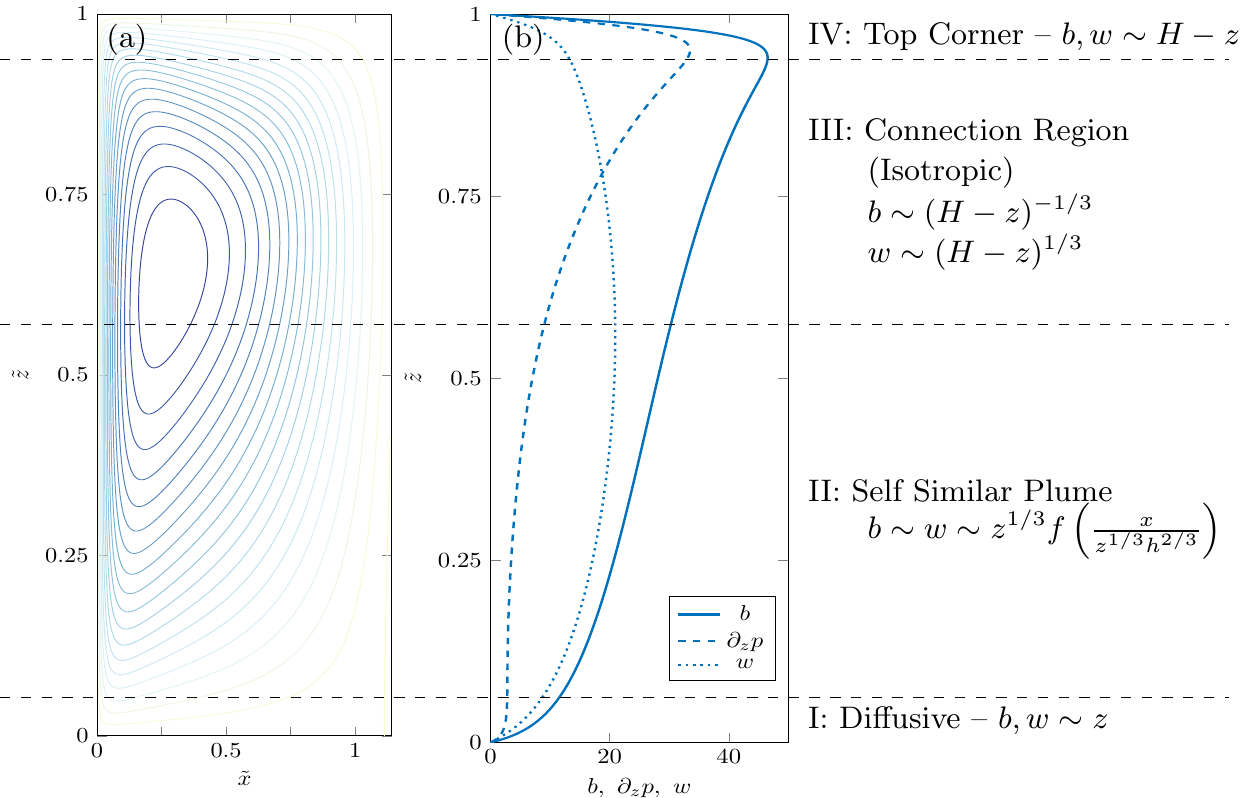}
			\caption{Regions of different behaviours observed for the flow close to the wall, with from left to right: contours of stream function in the numerical simulation; along wall profiles of $b$, $w$, and $\partial_z p$; and summary of the scalings predicted by the theory that are derived in \S~\ref{sec:results:theory}. DNS run at $Ra=200$ and $\Omega=1$.}
			\label{fig:roadmap}
		\end{figure}

	Figure~\ref{fig:roadmap} presents the global picture of the convective cell. We define four regions, labelled I through IV, based on the flow characteristics and their various scalings close to the wall, which are derived later. In the following subsections, we provide an analytical description of these different types of behaviours observed near the wall.
	
% -------------------------------------------------------------------------------- %
	\subsection{Theory}
	\label{sec:results:theory}
	
		Focusing on the $2$D dynamics in the plane $(x,z)$, we show that equations \eqref{eq2:16}, \eqref{eq2:17}, and \eqref{eq:advecdiffb}, all collapse into a single equation of a single function that will help us to understand the different physical scalings involved in the problem.
		
		From the continuity equation \eqref{eq2:16}, we define a stream function $\psi$ so that the horizontal and vertical velocities $u$ and $w$ are
		\begin{equation}
			u = - \dfrac{\partial \psi}{\partial z} \mathrm{~~~~~~~and~~~~~~~} w = \dfrac{\partial \psi}{\partial x}.
		\end{equation}
		Then, from the horizontal projection of Darcy's law, which involves cross-derivatives of the pressure field and of the stream function, we define a potential $\overline{\varphi}$ so that $p$ and $\psi$ can be written as
		\begin{equation}
			p = \dfrac{\partial\overline{\varphi}}{\partial z} \mathrm{~~~~~~~and~~~~~~~} \psi = \dfrac{\partial\overline{\varphi}}{\partial x}.
		\end{equation}
		Using this potential, the horizontal component of Darcy's law is automatically satisfied, and the vertical projection of Darcy's law becomes an equation for $\overline{\varphi}$ and the buoyancy $b$,
		\begin{equation}
			\mathbf{\nabla}^2 \overline{\varphi} = b.
		\end{equation}
		As a result, all relevant functions of the problem can be expressed using derivatives of the potential $\varphi$, as
	\begin{eqnarray}
		p &=& \partial_z \overline{\varphi}, \\
		\psi &=& \partial_x \overline{\varphi}, \\
		u &=& - \partial_x \partial_z \overline{\varphi}, \\
		w &=& \partial^2_x \overline{\varphi}, \\
		b &=& \mathbf{\nabla}^2 \overline{\varphi} = \partial^2_x \overline{\varphi} + \partial^2 _z \overline{\varphi}.
	\end{eqnarray}
		
		From these results and using the advection-diffusion equation for $b$~\eqref{eq:advecdiffb}, we obtain a collapsed equation for $\overline{\varphi}$ that is
		\begin{equation}
			\Omega \left[ \partial_t \mathbf{\nabla}^2 \overline{\varphi} - \partial_x\partial_z\overline{\varphi}~ \partial_x \mathbf{\nabla}^2 \overline{\varphi} + \partial^2_x\overline{\varphi}~ \partial_z \mathbf{\nabla}^2 \overline{\varphi} \right] = \mathbf{\nabla}^4 \overline{\varphi}.
		\end{equation}
		As $\Omega$ is a parameter of order $1$, by renormalising the potential $\overline{\varphi}$ and the time variable $t$ as
		\begin{equation}
			\overline{\varphi} ~\rightarrow~ \frac{1}{\Omega}\varphi \mathrm{~~~~~~~and~~~~~~~} t ~\rightarrow~ \Omega t,\label{eq:scale:varphi}
		\end{equation}
		we can write the equation for $\varphi$ without explicit dependence on $\Omega$. Hereafter we use $\varphi$ to represent the renormailised quantity according to~\eqref{eq:scale:varphi}. Hence, we have the following equation with three different contributions
		\begin{equation}
			\partial_t \mathbf{\nabla}^2 \varphi - \partial_x\partial_z\varphi \partial_x \mathbf{\nabla}^2 \varphi + \partial^2_x\varphi \partial_z \mathbf{\nabla}^2 \varphi = \mathbf{\nabla}^4 \varphi.
			\label{eq:potential}
		\end{equation}
		This equation~\eqref{eq:potential} represents a balance between temporal evolution of the system $\partial_t \mathbf{\nabla}^2 \varphi$, non-linearities coming from the advection term $\partial_x\partial_z\varphi \partial_x \mathbf{\nabla}^2 \varphi + \partial^2_x\varphi \partial_z \mathbf{\nabla}^2 \varphi$, and diffusion of buoyancy $\mathbf{\nabla}^4 \varphi$.
		
		In our study, we will consider steady states for buoyancy and velocity, so the temporal contribution will always be neglected, and relevant balances will involve the non-linear advection term and the diffusive term. We focus on steady solutions after the initial transient adjustment after warming begins (e.g. figure \ref{fig:psi}, left), but note that at sufficiently large Rayleigh number the transient dynamics can potentially trigger an instability, as shown in figure~\ref{fig:psi}(right). The two non-linear contributions in equation~\eqref{eq:potential} have the same scaling in terms of numbers of $x$ and $z$ derivatives, due to continuity.
		
		Note that the heat flux boundary condition~\eqref{eq:RaBC} can also be expressed in terms of $\varphi$ as
		\begin{equation}
			- \partial_x \mathbf{\nabla}^2 \varphi = Ra \Omega \mathrm{~~~~~at~~~~}x=0,
			\label{eq3:12}
		\end{equation}
		so that the only parameter group in the problem is the modified Rayleigh number $Ra \Omega$ which accounts for the effective heat capacity augmented by latent heat release or uptake from the phase change. 
		
		The buoyancy is given by the Laplacian of $\varphi$ and, as such, it involves $x$ and $z$ derivatives. These derivatives may or may not have the same scalings, which leads to three different idealised cases discussed in the following subsections:
		\begin{enumerate}
			\item If $\partial_x \sim \partial_z$, both derivatives contribute to the buoyancy and the region is isotropic.
			\item If $\partial_x \gg \partial_z$, the horizontal variations are more important than the vertical variations, which is characteristic of a vertical boundary layer.
			\item If $\partial_x \ll \partial_z$, the horizontal variations are less important than the vertical variations, which is characteristic of a horizontal boundary layer.
		\end{enumerate}
		
		We now examine the scalings resulting from equation~\eqref{eq:potential} based on these three regimes
		
% ................................................................................ %
		\subsubsection{Isotropic Scaling}
			\label{sec:IS}
		
			We define the following scalings for $x$, $z$, and $\varphi$
			\begin{equation}
				x,z ~\rightarrow~ L \mathrm{~~~~~~~and~~~~~~~} \varphi ~\rightarrow~ A,
			\end{equation}
			where $x$ and $z$ have the same scaling due to isotropy.
			
			The scaling of the heat flux boundary condition leads to
			\begin{equation}
				A \sim Ra \Omega L^3 \sim L^3.
				\label{eq3:15}
			\end{equation}			
			We assume that the system is in a steady state, so that temporal derivatives have no contribution, the scaling of the remaining terms in the potential equation~\eqref{eq:potential} yields
			\begin{eqnarray}
				-\partial_x\partial_z\varphi \partial_x \mathbf{\nabla}^2 \varphi + \partial^2_x\varphi \partial_z \mathbf{\nabla}^2 \varphi & \sim & \frac{A^2}{L^5} ~\sim~ Ra^2 \Omega^2 L, \\
				\mathbf{\nabla}^4 \varphi & \sim & \frac{A}{L^4} ~\sim~ \frac{Ra \Omega}{L},
			\end{eqnarray}
			where we eliminate $A$ using~\eqref{eq3:15}. This means that the non-linear terms and the diffusive term have different scalings, in $L$ and in $1/L$ respectively. Two asymptotic behaviours can therefore be described:
			\begin{enumerate}
				\item $L \ll (Ra\Omega)^{-1/2}$: The diffusive term dominates the equation and is at least one order of magnitude in $L$ larger than the non-linear terms. The regime is diffusive, which means that $\mathbf{u} \cdot \nabla b \approx 0$, and the buoyancy satisfies a Poisson equation with heat source from the heated wall. From the scaling of $\varphi$ in~\eqref{eq3:15}, we have $b\sim Ra \Omega L$ and $w\sim Ra \Omega L$;
				\item $L \gg (Ra\Omega)^{-1/2}$: The non-linear terms dominate the equation.
			\end{enumerate}
		
% ................................................................................ %
		\subsubsection{Isotropic Stagnation Flow Scaling}
		
			Another explaination for a linear scaling comes from a stagnation point flow model. We illustrate this here for flow near the upper boundary at $z=H$. A Taylor series expansion of the stream function in to second order gives
			\begin{equation}
				\psi = A x + B z + C x^2 + D x z + E z^2,
			\end{equation}
			Enforcing boundary conditions that $w=\partial_x \psi=0$ at $z=H$ and the symmetry condition $u=-\partial_z \psi=0$ at $x=0$ yields the following stream function
			\begin{equation}
				\psi = -D x (H-z),
			\end{equation}
			with $D$ a coefficient to be determined. This suggests a linear scaling in $z$ for $w$ near to the upper boundary. The coefficient $D$ depends on matching to the flow from the incoming isotropic region; this calculation is challenging and not pursued here. A similar analysis yields a corresponding expression $\psi \propto xz$ near the lower boundary at $z=0$.
		
% ................................................................................ %
		\subsubsection{Vertical Boundary Layer}
			\label{sec:VBL}
		
			We define the following scalings for $x$, $z$, and $\varphi$
			\begin{equation}
				x ~\rightarrow~ L_x, \mathrm{~~~~~~~} z ~\rightarrow~ L_z \sim z, \mathrm{~~~~~~~and~~~~~~~} \varphi ~\rightarrow~ A,
			\end{equation}
			where $L_z \sim z$. In this case, the variations along $z$ are small compared to the variations along $x$ and the scaling length $L_x$ is expected to be a function of $z$ (c.f. similarity solutions of \cite{cheng1977, guba2006}).
		
			We assume that the system is in a steady state, so that temporal derivatives have no contribution. Given that $\partial_x \gg \partial_z$, we neglect the $z$ derivatives compared to $x$ derivatives and the scaling of the potential equation is given as
			\begin{equation}
				\textcolor{black!30}{\partial_t \mathbf{\nabla}^2 \varphi} - \underbrace{\partial_x\partial_z\varphi \partial_x (\partial^2_x \varphi \textcolor{black!30}{+ \partial^2_z \varphi}) + \partial^2_x\varphi \partial_z (\partial^2_x \varphi \textcolor{black!30}{+ \partial^2_z \varphi})}_{\frac{A^2}{L_z L_x^4}} = \underbrace{\partial^4_x \varphi}_{\frac{A}{L_x^4}} \textcolor{black!30}{+ \partial^4_z \varphi},
				\label{eq3:20}
			\end{equation}
			leading to the balance
			\begin{equation}
				A \sim L_z \sim z.
			\end{equation}
			
			The same method gives a scaling of the wall heat flux boundary condition~\eqref{eq3:12}
			\begin{equation}
				Ra\Omega  \simeq - \underbrace{\partial_x (\partial^2_x \varphi}_{\frac{A}{L_x^3}} \textcolor{black!30}{+ \partial^2_z \varphi}),
			\end{equation}
			then
			\begin{equation}
				L_x \sim \left(\frac{A}{Ra\Omega}\right) ^{1/3} \propto z^{1/3}.
			\end{equation}
			
			From these scalings, we define a characteristic boundary layer width $L_x$ as
			\begin{equation}
				L_x = \frac{h^{2/3} z ^{1/3}}{\Omega^{1/3}} = z^{1/3} (Ra \Omega)^{-1/3}
			\end{equation}
			where $h$ is the dimensionless characteristic length defined from the Rayleigh number and the size of the wall $H$ in equation~\eqref{eq:h}. This motivates using a self-similar variable
			\begin{equation}
				\eta = \frac{x}{L_x} = \frac{x \Omega^{1/3}}{h^{2/3} z^{1/3}}.
			\end{equation}
			The potential can be written under the form $\varphi = z f(\eta)$ in terms of a function $f$. As we are neglecting the $z$ derivatives in the boundary layer, we have
			\begin{equation}
				w = b = \Omega^{-1} \partial^2 _ x \varphi \mathrm{~~~~~~~and~~~~~~~} p = \Omega^{-1} \partial_z \varphi,
			\end{equation}
			and we obtain the following scalings for $w$ and $b$
			\begin{equation}
				w \sim Ra ^{2/3} \Omega^{-1/3} z^{1/3} \mathrm{~~~~~~~and~~~~~~~} b \sim Ra ^{2/3} \Omega^{-1/3} z^{1/3}.
				\label{eq:scalingsR2}
			\end{equation}
			These scalings are analagous to those derived by \cite{cheng1977} for a self similar flow at a wall with imposed temperature varying $T\sim z^{1/3}$, which recovers a constant heat flux. Replacing $\varphi = z f(\eta)$ in the collapsed advection-diffusion equation~\eqref{eq3:20}, we derive a self-similar ordinary differential equation for $f$,
			\begin{equation}
				(f'')^2 - 2 f' f''' - 3 f'''' = 0,
				\label{eq:selfsimilar2}
			\end{equation}
			The boundary conditions imposed at the wall ($x=0$) and in the far-field (see similar analysis in \cite{cheng1977} for a boundary layer in a porous medium with wall temperature varying as a power law of distance along the wall, and \cite{guba2006} for a boundary layer next to an isothermal boundary in a solidifying mush layer) are
			\begin{equation}
				\underbrace{f'''(0) = -1}_{\mathrm{flux~at~}x=0}, \mathrm{~~~} \underbrace{f'(0) = 0}_{u=0\mathrm{~at~the~wall}} \mathrm{~~~and~~~} \underbrace{f''(\eta^*) \rightarrow 0 \mathrm{~~~as~~~} \eta^* \rightarrow \infty }_{\mathrm{constant~}b\mathrm{~in~the~far~field}}.
				\label{eq:selfsimilar2BC}
			\end{equation}
			Note that the final condition of constant buoyancy in the far field is equivalent to zero vertical velocity in the far field outside of the boundary layer. For numerical integration purposes, we set $f''(\eta^*) = 0$ at some large value of $\eta^*$ and thereafter check that the solution becomes independent of this choice of $\eta^*$. Although~\eqref{eq:selfsimilar2} is a fourth-order differential equation, only three boundary conditions are required because the physical variables only depend on derivatives of $f$ and~\eqref{eq:selfsimilar2} is a third order equation for $f'$. We later compute this solution using the Dedalus iterative boundary value problem solver \citep{burns2019} over $256$ Chebyshev nodes, with $\eta^{\star}$ chosen large enough to ensure that the far field boundary condition is satisfied. Numerical convergence of the solution is obtained after a few iterations.
		
% ................................................................................ %
		\subsubsection{Horizontal Boundary Layer}
			\label{sec:HBL}
		
			To determine scalings for the horizontal boundary layer near the top boundary, we define the following scalings for $x$, $z$, and $\varphi$
			\begin{equation}
				x ~\rightarrow~ L_x \sim x, \mathrm{~~~~~~~} z ~\rightarrow~ L_z, \mathrm{~~~~~~~and~~~~~~~} \varphi ~\rightarrow~ A,
			\end{equation}
			where $L_x \sim x$ and $L_z \ll L_x$ as the variations along $x$ are small compared to the variations along $z$. The scaling length $L_z$ is expected to be a function of $x$.
		
			We again assume that the system is in a steady state, so that temporal derivatives have no contribution. Given that $\partial_x \ll \partial_z$, we neglect the $x$ derivatives and the scaling of the potential equation yields
			\begin{equation}
				\textcolor{black!30}{\partial_t \mathbf{\nabla}^2 \varphi} - \underbrace{\partial_x\partial_z\varphi \partial_x (\textcolor{black!30}{\partial^2_x \varphi +} \partial^2_z \varphi) + \partial^2_x\varphi \partial_z (\textcolor{black!30}{\partial^2_x \varphi +} \partial^2_z \varphi)}_{\frac{A^2}{L_z^3 L_x^2}} = \textcolor{black!30}{\partial^4_x \varphi +} \underbrace{\partial^4_z \varphi}_{\frac{A}{L_z^4}}.
			\end{equation}
			Balancing these terms leads to
			\begin{equation}
				A \sim L_x^2 / L_z \sim x^2 / L_z,
			\end{equation}
			where $L_z(x)$ has to be determined.
			
			As for the vertical boundary layer, we introduce a similarity variable $\zeta$ and write the potential $\varphi$ in terms of a function $g$ as
			\begin{equation}
				\zeta = \frac{z}{L_z(x)} \mathrm{~~~~~~~and~~~~~~~} \varphi = \frac{x^2}{L_z(x)} g (\zeta).
			\end{equation}
			
			We determine $L_z (x)$ by assuming that an order $1$ fraction of the buoyant heat flux coming from below towards the top boundary is eventually transported sideways through this upper boundary layer, so that the horizontal spreading flow carries a constant heat flux at leading order. A full solution of the problem will require asymptotic matching of the incoming fluxes. However, to determine the scalings this only needs to be true in an order-of-magnitude sense; the whole buoyant heat flux, corresponding to the heat flux at the entire wall (i.e. sum of the flux $-b_x = Ra= 1/h^2$ emitted over the wall of length $H$), is going sideways through this top region, meaning
			\begin{equation}
				\int u b \diff z \sim \frac{Ra}{h^2}\mathrm{~~~~~~~as~~~~~~~} h \rightarrow 0,
				\label{eq3:34}
			\end{equation}
			where the integral is over the depth of the horizontal boundary layer. This assumes the heat lost by conduction through the top boundary does not change the order of the magnitude of the heat flux: this approximation should work well for small enough $x$ sufficiently close to the plume, but may break down sufficiently far downstream where the accumulated heat loss may eventually become substantial. Noting that $u=-\partial_x\partial_z\varphi/\Omega$ and $b \approx \partial_z^2 \varphi/\Omega$ within the boundary layer, computing the left-hand side yields
			\begin{equation}
				\int u b \diff z = -\frac{1}{\Omega^2}\frac{x^3}{L_z (x) ^4} \int \left[ g'' (\zeta) \left( 2 g'(\zeta) - \frac{x L'_z (x)}{L_z(x)} \left( \zeta g'' (\zeta) + 2 g' (\zeta) \right)\right) \right] \diff \zeta.
			\end{equation}
			If $L_z$ has power-law dependence on $x$, then $x L_z'/L_z$ is a constant. Then, the balance of the heat fluxes \eqref{eq3:34} gives
			\begin{equation}
				L_z (x) = \frac{h^{1/2} x ^{3/4}}{H^{1/4} \Omega^{1/2}} = x^{3/4} Ra^{-1/4} \Omega^{-1/2}
,
			\end{equation}
			where, we retain a dimensionless $H$ in the following calculation to aid interpretation, even though the dimensionless height is $1$. Hence, the potential $\varphi$ is given as
			\begin{equation}
				\varphi = \frac{\Omega^{1/2} H^{1/4} x^{5/4}}{h^{1/2}}g (\zeta) = Ra^{1/4} \Omega^{1/2} H^{1/4}x^{5/4} g(\zeta).
			\end{equation}
			As we are neglecting the $x$ derivatives, we have
			\begin{equation}
				w = \partial^2_x \varphi /\Omega \mathrm{~~~~~~~and~~~~~~~} b \simeq \partial_z p = \partial_z ^2 \varphi /\Omega,
			\end{equation}
			which implies that hydrostatic balance holds at leading order in this boundary layer, and the pressure gradient is
			\begin{equation}
				\partial_z p = \partial_z ^2 \varphi/\Omega = \frac{\Omega^{1/2} H^{3/4}}{h^{3/2} x^{1/4}} g'' (\zeta) = Ra^{3/4} \frac{\Omega^{1/2} H^{3/4}}{x^{1/4}} g''(\zeta).
				\label{eq:pressureR4}
			\end{equation}

% -------------------------------------------------------------------------------- %
	\subsection{Comparison to numerical solutions}

		Using the above scaling arguments, we can define four different domains for the numerical simulation results presented in figure~\ref{fig:roadmap}, labeled I, II, III, and IV, starting from the bottom of the wall, and correspond to the different scalings derived in the previous subsections. These domains yield contrasting behaviour of the along wall profiles of $w$, $b$, and $\partial_z p$ (see figure~\ref{fig:roadmap}). The interest in these three quantities is motivated by the relation $b = \mathbf{\nabla}^2 \varphi = w + \partial_z p$, showing that the buoyancy, the vertical velocity, and the vertical pressure gradient, are linearly related.
		
		These four regions have different scalings, that can be explained as follows.
		
		\begin{enumerate}
			\item[\textbf{Region I}] Tip behaviour, in a small regularisation region. All quantities are small and the region is isotropic. Diffusive terms dominate and the buoyancy approximately satisfies a Laplace equation for diffusive heat transfer in steady state. Because this region is isotropic, we expect the variation along the length of the wall to satisfy $w\sim z$ and $b\sim z$ as discussed in section~\ref{sec:IS};
			\item[\textbf{Region II}] Rising plume region, described by a vertical boundary layer in which horizontal spatial derivatives dominate. The vertical Darcy flow is driven by buoyancy forces with pressure gradients negligible, yielding $b \simeq w$. Due to the vertical boundary layer, $z^{1/3}$ scalings are expected as discussed in section~\ref{sec:VBL};
			\item[\textbf{Region III}] Connection region, between II and IV, where $b$, $w$, and $\partial_z p$ are of the same order of magnitude. The region is isotropic and the non-linear terms dominate the collapsed equation. The incoming plume flow from below carries most of the buoyancy flux, which is advected through this region and the additional heating at the wall does not change the buoyancy flux substantially, remaining of the same order of magnitude throughout this region. An empirical power law of $w \propto (H-z)^{1/3}$ and $b \propto (H-z)^{-1/3}$ scalings is discussed in section~\ref{sec:reg3}.
			\item[\textbf{Region IV}] Top corner region, with the scaling of a top boundary layer, bringing the heat flux sideways into the horizontal boundary layer. Again, this region is diffusive and we expect the linear scalings, similar to those derived in section~\ref{sec:IS} except using the distance to $(0,H)$. Region IV forms part of the turnaround region, where the initially vertical flow adjusts into horizontal flow towards a region that grows with $x$ with a horizontal boundary layer as in section~\ref{sec:HBL} and its scalings for the pressure gradient. Region IV can be decomposed into two distinct region, with a corner region close to the top of the wall, which develops into an adjacent region as a horizontal top boundary layer. We hypothesise that the pressure gradient from the horizontal boundary layer is imprinted on the corner region, leading to a unique pressure scaling for region IV.
		\end{enumerate}
		
		\begin{figure}
			\centering
			\includegraphics[scale=1]{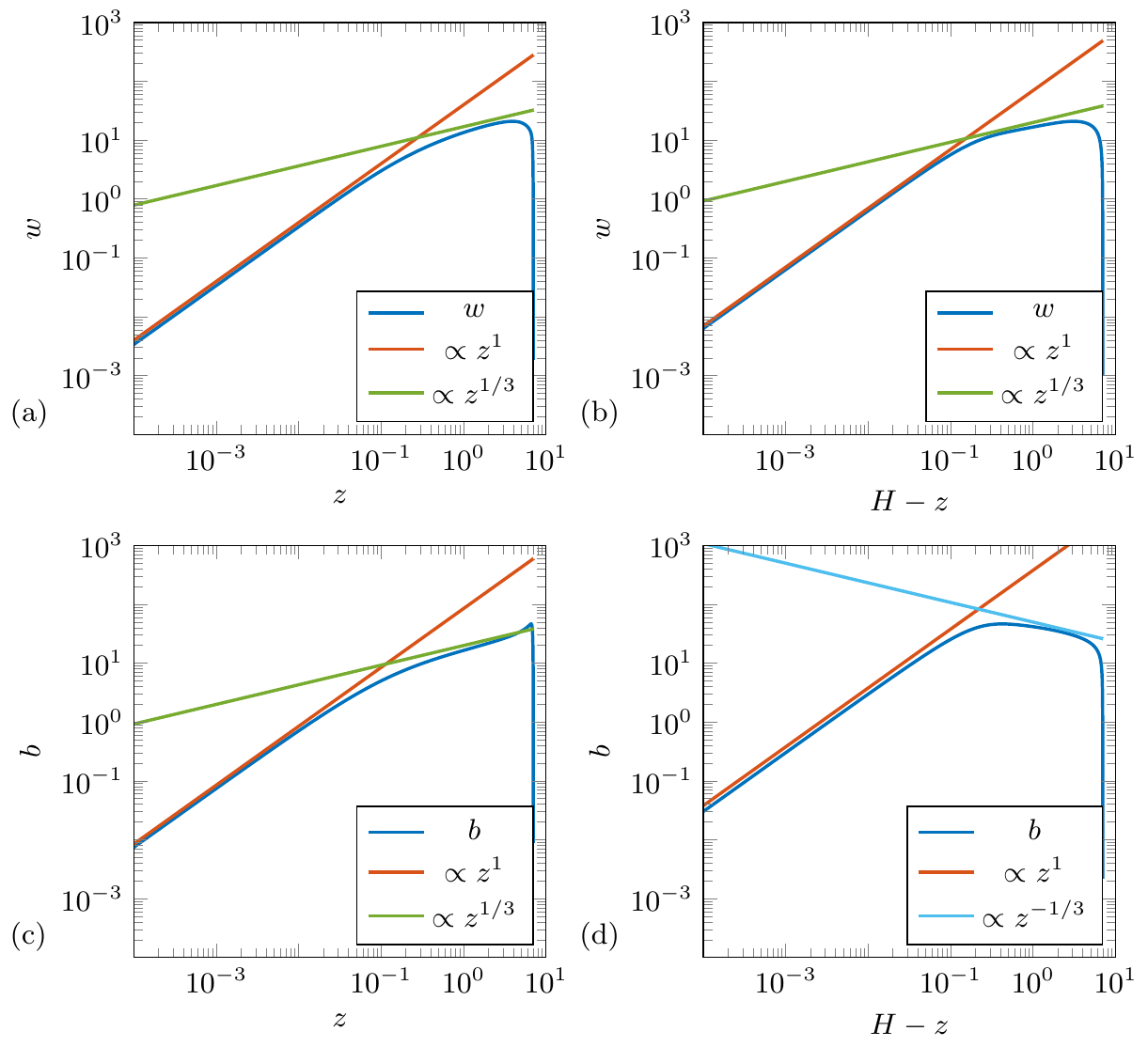}
			\caption{Scalings infered from the profiles along the wall. Blue lines show (a): Vertical velocity varying with distance from the bottom; (b): Vertical velocity varying with distance from the top; (c): Buoyancy varying with distance from the bottom; and (d): Buoyancy varying with distance from the top. Red lines indicate slope of linear scalings, green lines for $+1/3$ power law, and cyan lines for $-1/3$ power law.} % \textcolor{red}{do we need to go to such small $z$ on these plots? I wonder if we might enhance the comparison of the $z^{1/3}$ scalings by clipping the axes at $log(z)\sim 10^{-3}$  or $log(z)\sim 10^{-4}$? Perhaps worth a try and check? }
			\label{fig:loglog}
		\end{figure}		
		
		In figure~\ref{fig:loglog}, we present $\log$-$\log$ plots of the vertical velocity and of the buoyancy in which the different scalings predicted analytically in the previous subsections are tested. Figures~\ref{fig:loglog}(a) and (b) show $\log$-$\log$ plots of the vertical velocity along the wall, with distance from the bottom and distance from the top, respectively. We first observe a linear scaling $w\propto z$ in region I, behaviour potentially consistent with two $1/3$ scalings $w\propto z^{1/3}$ and $w\propto (H-z)^{1/3}$ in region II and III, and a linear scaling $w\propto (H-z)$ in region IV. Figures~\ref{fig:loglog}(c) and (d) show $\log$-$\log$ plots of the buoyancy varying with distance along the wall, starting from the bottom and starting from the top, respectively. Similarly to the vertical velocity, we first observe a linear scaling $w\propto z$ in region I, a $1/3$ scalings $w\propto z^{1/3}$ in region II and a $-1/3$ scaling $w\propto (H-z)^{-1/3}$ in region III, and a linear scaling $w\propto (H-z)$ in region IV. As discussed above, scalings for regions I, II, and IV, are consistent with the theoretical expectations. Region III, however, shows interesting empirically suggested scalings for which we do not yet have an asymptotic explanation.
		We now compare our scaling predictions with the numerics in more detail.

% ................................................................................ %
		\subsubsection{Region I}
	
		Close to the bottom tip, the vertical velocity, the buoyancy, and  the pressure gradient are small (figure~\ref{fig:roadmap}). All terms contribute to the equations as the region is isotropic. The scalings described in section~\ref{sec:IS} show that, because the length scale is small, this region is diffusive and has a linear scaling for $b$ and $w$ with $z$, that is observed in figure~\ref{fig:loglog}(a) and (c). This region breaks down when the isotropic scale $L_x \sim z$ becomes comparable with the width $L_x \sim h^{2/3} z^{1/3}$ of a potential rising plume, i.e. when $z$ is of order $h$.

% ................................................................................ %
		\subsubsection{Region II}

\begin{figure}
			\centering
			\includegraphics[scale=1]{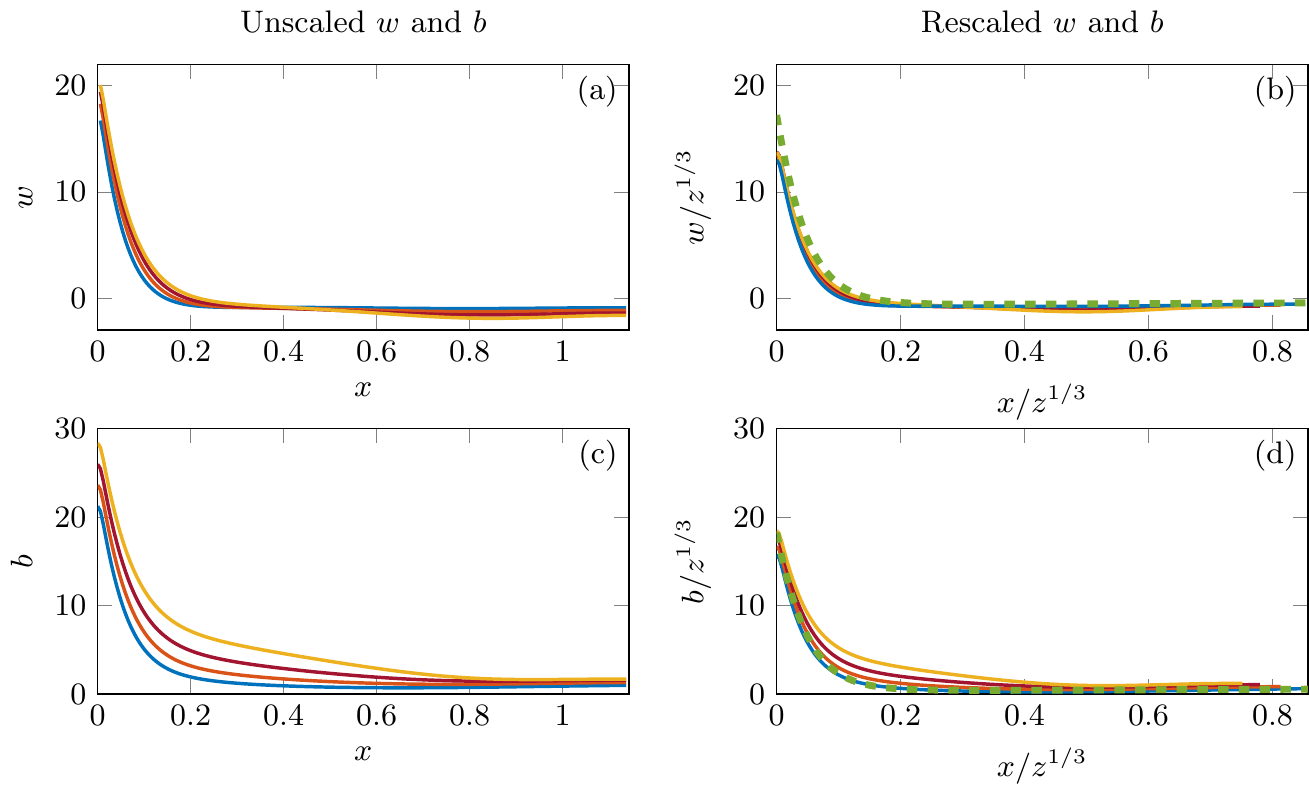}
			\caption{Horizontal slices of vertical velocity $w$ and buoyancy $b$ at different depths in region II (at $z=0.34$ (blue), $z=0.39$ (orange), $z=0.44$ (red), and $z=0.49$ (yellow)), with: (a) $w$ vs $x$ for different $z$; (b) collapsed $w$ vs collapsed $x$; (c) $b$ vs $x$ for different $z$; and (d) collapsed $b$ vs collapsed $x$. The self-similar solkution to \eqref{eq:selfsimilar2}-\eqref{eq:selfsimilar2BC} is superimposed to the rescaled profiles in figures (b) and (d) (dashed line).} % depth: 130/384, 150/384, 170/384, 190/384
			\label{fig:R2}
		\end{figure}

		In region II, the flow acts like a rising buoyant plume described by the vertical boundary layer from section~\ref{sec:VBL}. The flow structure is similar to the analysis performed by \cite{guba2006} but with some differences in the power law scalings in the similarity solution: where \cite{guba2006} derived a self-similar variable in $z^{-1/2}$ and velocity and buoyancy scalings in $z^{1/2}$ for an isothermal wall, we here derive solutions for a constant flux wall with a self-similar variable in $z^{-1/3}$ yielding velocity and buoyancy scalings in $z^{1/3}$. Note that these scales match the \cite{cheng1977} solution with $\lambda=1/3$. By equating the lengthscales $L_x$ between region I and region II, we see that the lateral extent of this region at the bottom, connecting to region I, is initially of order $h$, and thereafter increases proportional to $h^{2/3} z^{1/3}$. From equation~\eqref{eq:scalingsR2}, our theory predicts a scaling in $z^{1/3}$ for $w$ and $b$ with the numerical solutions appearing to approach this scaling for intermediate $z$ in the $\log$-$\log$ plot in figures~\ref{fig:loglog}(a) and (c). In figure~\ref{fig:R2}, we present horizontal slices of the vertical velocity (top) and of the buoyancy (bottom) at different depths in region II (plain lines), rescaled in the horizontal variable by $z^{1/3}$ as well as in amplitude. We observe is a good collapse at $x=0$, close to the wall. The superimposed dashed line is the self-similar solution of equation~\eqref{eq:selfsimilar2} and~\eqref{eq:selfsimilar2BC}. Note that the scale separation between the start $z=O(h)$ and end $z=O(1)$ of this region is quite modest for the given Rayleigh number $Ra=200$, and there is only around a decade for the power law scaling to adjust. Whilst the scale separation would increase for a steady laminar flow at higher $Ra$, our simulations (figure~\ref{fig:psi}(b)) revealed that instability and unsteady flow develops at larger $Ra$.
		
			The computed solutions show a good agreement with the data. The profiles diverge from the similarity solution in the far field, due to the influence of the return flow and its impact on the background buoyancy field that is not included in the boundary layer model. Nevertheless, the similarity variables yield a good collapse of the shape of decay of the velocity and buoyancy profiles away from the heat source, their characteristic boundary layer width and capture their leading order quantitative magnitude. We note that the peak in $w$ measured from the DNS is not as large as predicted by the similarity solution: this might be due to the neglect of the pressure gradient term in the leading order balance for the similarity solution, which acts to slightly decelerate the flow in the full numerical solution.

% ................................................................................ %
		\subsubsection{Region III}
		\label{sec:reg3}
		
		Region III is an isotropic region connecting regions II and IV. Figure~\ref{fig:roadmap}(b) shows how the force balance in this connection region behaves. We recall that $b = w + \partial_z p$	from Darcy's law~\eqref{eq2:16}. In region II, as seen before, the vertical boundary layer model leads to $w \simeq b$ and $\partial_z p$ is negligible, whereas in region IV, $b \simeq \partial_z p$ and $w$ has a small contribution (see figure~\ref{fig:roadmap}(b)). The only way this configuration can occur is to have an isotropic transition region in between, in which $w$ and $\partial_z p$ have the same order of magnitude, with $w$ decreasing while $\partial_z p$ increases. This behaviour is observed in domain III in figure~\ref{fig:roadmap}(b).
		
		\begin{figure}
			\centering
			\includegraphics[scale=1]{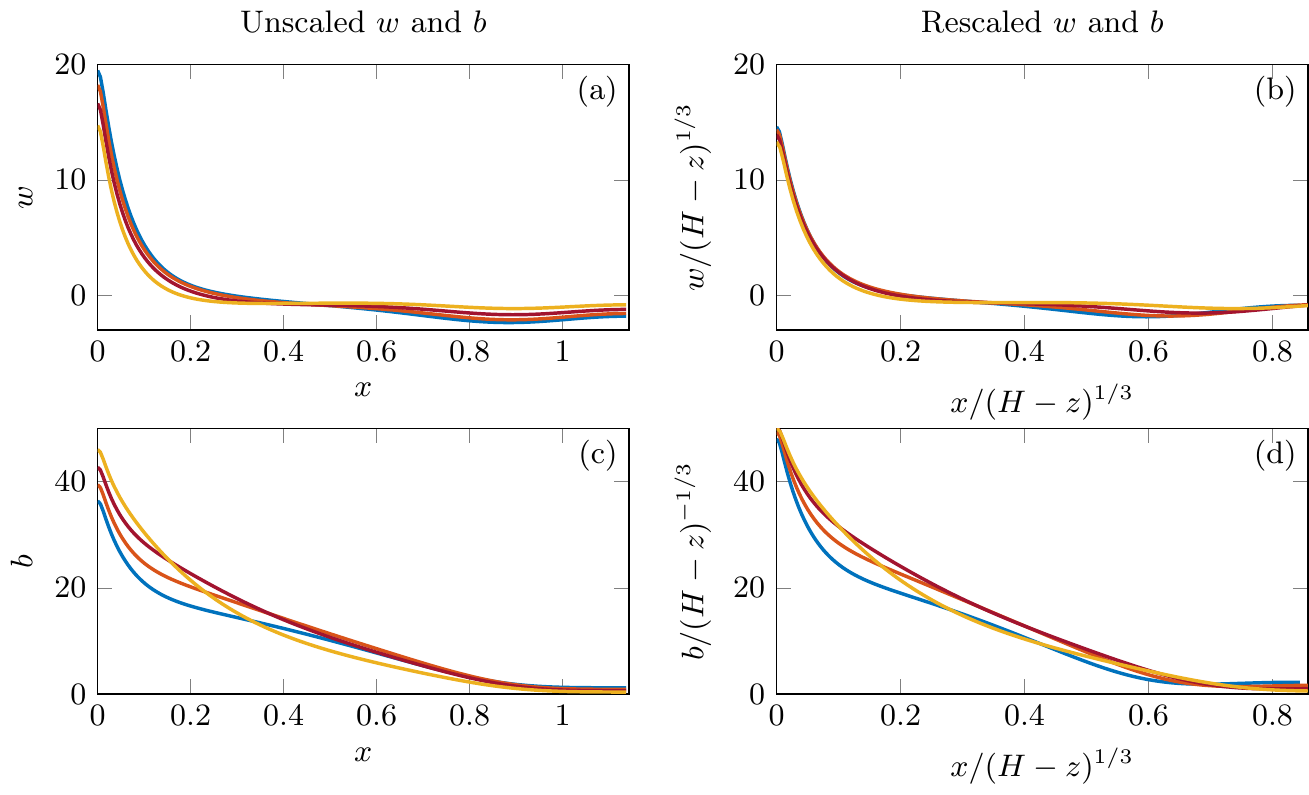}
			\caption{Horizontal slices of vertical velocity $w$ and buoyancy $b$ at different depths in region III (at $z=0.65$ (blue), $z=0.70$ (orange), $z=0.75$ (red), and $z=0.80$ (yellow)), with: (a) $w$ vs $x$ for different $z$; (b) collapsed $w$ vs collapsed $x$; (c) $b$ vs $x$ for different $z$; and (d) collapsed $b$ vs collapsed $x$.} % Depth: 250/384, 270/384, 290/384, 310/384
			\label{fig:R3}
		\end{figure}

		From the $\log$-$\log$ plots in figures~\ref{fig:loglog} (b) and (d), we infer two empirical scalings: a $(z-H)^{1/3}$ scaling for $w$, and a $(z-H)^{-1/3}$ scaling for $b$. Figure~\ref{fig:R3} presents horizontal slices of vertical velocity (top) and buoyancy (bottom) rescaled by $1/3$ and $-1/3$ power laws, showing a good collapse close to the wall. A deviation in the far field, due to the return flow outside of the boundary layer, is observed. The collapse is better for $w$ than for $b$, pointing towards a transition between a vertical boundary layer in which $w$ is set and a fully isotropic region dominated by fluxes. This collapse, however, is curious as it does not correspond to the boundary layer described in the previous sections, and still awaits a theoretical explanation which we reserve for future work. Note that this scaling implies that the product $wb$ is approximately constant, which means that the heat flux advected through this region III also remains approximately constant.

% ................................................................................ %
		\subsubsection{Region IV}
		
		In the top corner region IV, both the scalings of $b$ and $w$, linear with $H-z$ (see figure~\ref{fig:loglog}), are consistent with a stagnation point flow. The velocity profile decelerates to satisfy the non-normal flow boundary condition at the top, and we observe the dominance of diffusive transfers of buoyancy as the velocity decays to zero in the corner.

\begin{figure}
			\centering
			\includegraphics[scale=1]{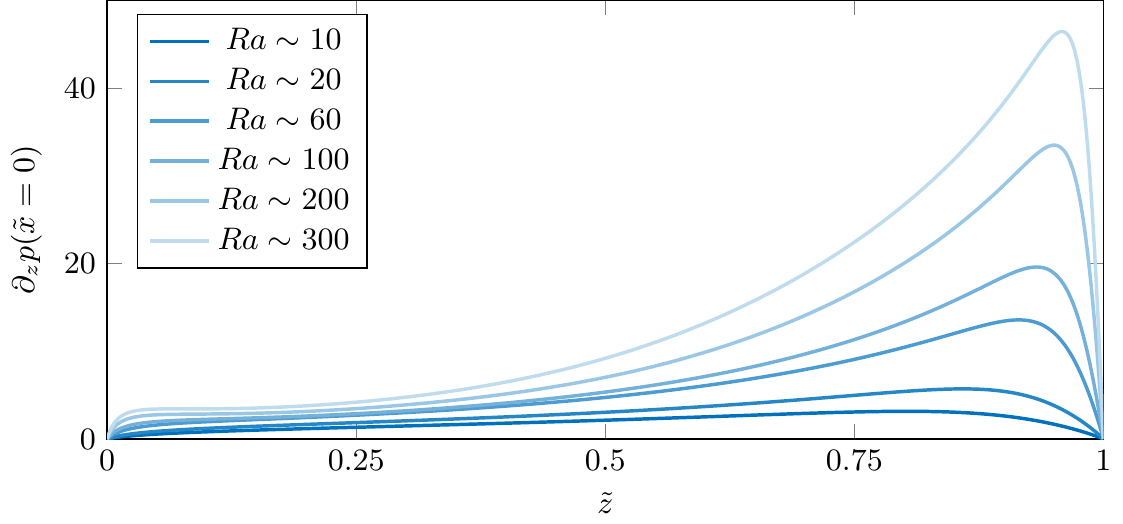}
			\caption{Along wall profiles of the vertical pressure gradient for different Rayleigh numbers (corresponding to different heating prefactors $Q_0$).}
			\label{fig:pcrack}
		\end{figure}

\begin{figure}
			\centering
			\includegraphics[scale=1]{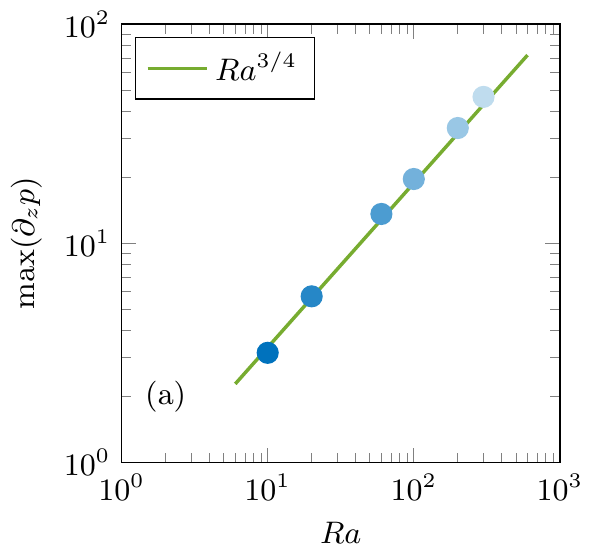}
			\includegraphics[scale=1]{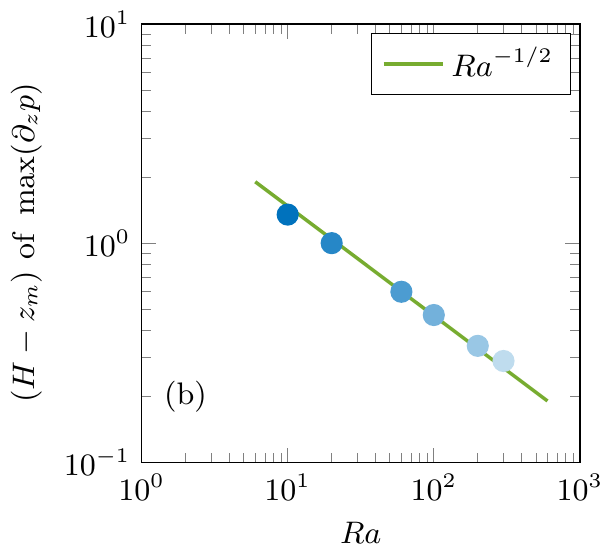}
			\caption{Scaling laws for along wall pressure with Rayleigh number in $\log$-$\log$ plots. The shades of blue correspond to the Rayleigh numbers indicated in figure~\ref{fig:pcrack}. (a) Amplitude of the peak of pressure gradient, with a $Ra^{3/4}$ law. (b) Location of the maximum of the pressure gradient, as a distance from the top boundary, with a $Ra^{-1/2}$ law.}
			\label{fig:logpcrack}
		\end{figure}

		We also note that the pressure gradient is consistent with the imprint of scalings in the neighbouring horizontal boundary layer in which the flow spreads horizontally. In this region, the pressure gradient reaches a maximum and balances the buoyancy force at leading order. According to equation~\eqref{eq:pressureR4}, the peak in the vertical profile of the pressure gradient should increase and move upwards as we increase the Rayleigh number. Figure~\ref{fig:pcrack} shows results from DNS at various Rayleigh numbers, with this increasing peak. For each of these profiles, we measure the amplitude of the peak $\max (\partial_z p)$ and its location $z_m$ from the top of the wall. Figure~\ref{fig:logpcrack} presents these data in $\log$-$\log$ plots. For the amplitude of the peak (figure~\ref{fig:logpcrack}(a)) we have a $Ra^{3/4}$ scaling (or $h^{-3/2}$) consistent with the theory developped in section~\ref{sec:HBL}. For the location $z_m$ (figure~\ref{fig:logpcrack}(b)), we have a $Ra^{-1/2}$ scaling (or $h$) for $H-z_m$, which is linked to the crossover lengthscale derived in section~\ref{sec:IS} where both the nonlinear advective and diffusive terms become of similar magnitude. Note that the testing of these scalings in figure~\ref{fig:logpcrack} is limited in the range of accessible Rayleigh numbers: at low $Ra$, the flow stops convecting and there is no boundary layers in which our model would apply; conversely, at high $Ra$, the near-crack instability is triggered (see figure~\ref{fig:psi}) and the theoretical derivation does not hold anymore.
		
\begin{figure}
			\centering
			\includegraphics[scale=1]{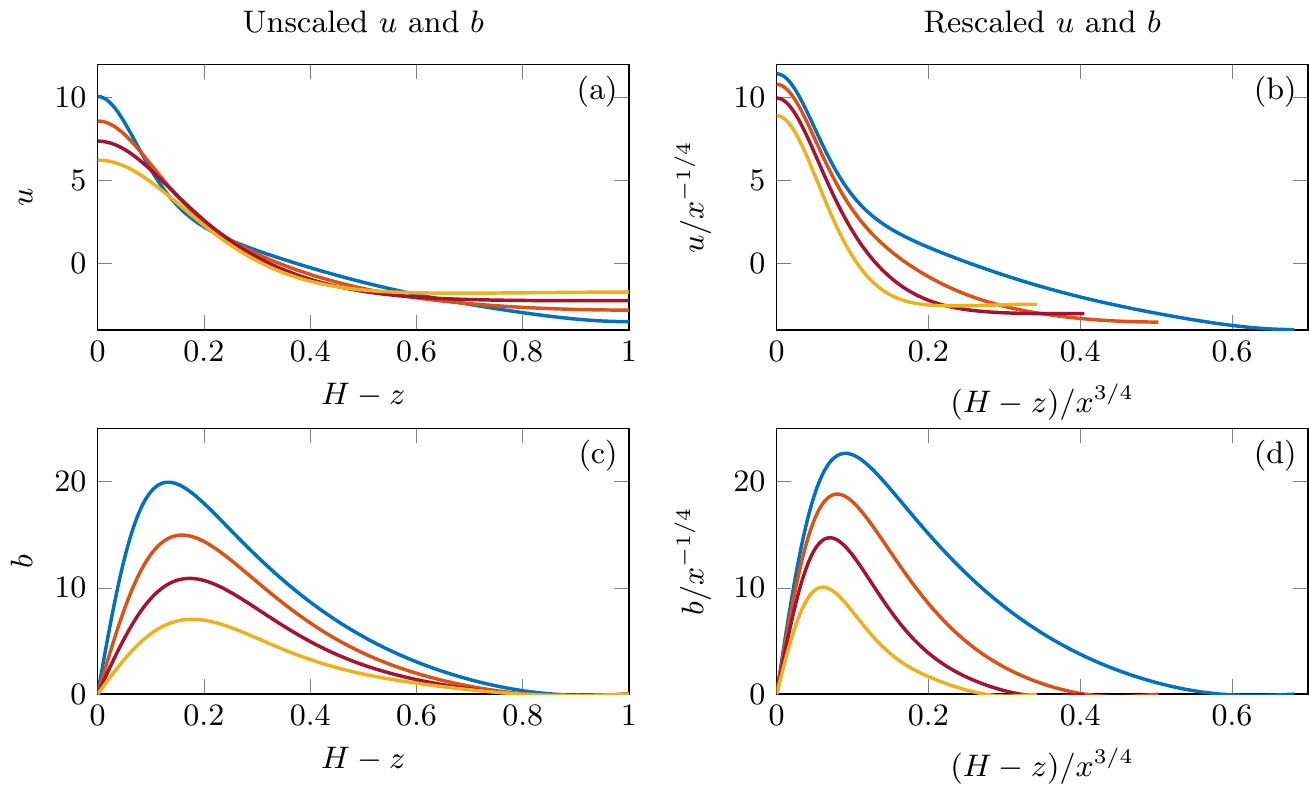}
			\caption{Vertical slices of horizontal velocity $u$ and buoyancy $b$ at different locations $x$ in region IV (mesured from the crack: at $z=0.24$ (blue), $z=0.36$ (orange), $z=0.48$ (red), and $z=0.60$ (yellow)), with: (a) $u$ vs $(H-z)$ for different $x$; (b) collapsed $u$ vs collapsed $(H-z)$; (c) $b$ vs $(H-z)$ for different $x$; and (d) collapsed $b$ vs collapsed $(H-z)$.} % locations in pixels x=192+(40,60,80,100)
			\label{fig:HBLcollapse}
		\end{figure}

		We present, in figure~\ref{fig:HBLcollapse}, vertical slices of the horizontal velocity $u$ and of the buoyancy $b$ unscaled (left) and rescaled by the scalings predicted by the theory (right). The self-similar scalings works relatively well in a thin boundary layer close to the top boundary $z=H$ but breaks further below. It is likely that there is some more complex behaviour going on with the spreading horizontal boundary layer fed by the rising plume, with another diffusive boundary layer nested within this due to the cooling boundary condition that $b=0$ at the top of the domain.
		

% -------------------------------------------------------------------------------- %
	\subsection{Prediction of the Solid Dissolution Rate}
		
		We now discuss the phase change behaviour close to the wall. We have previously shown that the flow transports solute which can drive either local dissolution of the ice matrix, or freezing, which modify the solid fraction.
		
		The advection-diffusion equation for $b$ and the solute and heat equations give an equation for the evolution of the solid fraction $\phi$, or melting rate, depending on the potential $\varphi$. At first order, the solid fraction can be written $\phi = \overline{\phi} + \hat{\phi}$, where $\overline{\phi}$ is the initial value and $\hat{\phi}$ the perturbation away from the steady state. Recalling that the near-eutectic approximation lead to $b(1-\phi)\approx b$, then (2.15) yields an equation for $\partial \phi/\partial t$ in terms of $b$. Substituting for $\partial b/\partial t+\mathbf{u}\cdot \nabla b$ using \ref{eq:advecdiffb} and noting that $b = \nabla^2 \varphi/\Omega $, in dimensionless form this yields
		\begin{equation}
			\dfrac{\partial \hat{\phi}}{\partial t} = -\frac{1}{\mathscr{C} Ra} \left[ \frac{\partial b}{\partial t} + \mathbf{u} \cdot \nabla b\right] = -\frac{1}{\mathscr{C} Ra \Omega} \mathbf{\nabla}^4 \varphi.
			\label{eq:melting}
		\end{equation}
		The intermediate expression in equation~\eqref{eq:melting} indicates the role of the advection of salinity anomalies, and hence buoyancy anomalies, in controlling the phase change. There is a tendency for reduction in solid fraction by local dissolution when $\mathbf{u} \cdot \nabla b >0$,  which corresponds to an advection of saline fluid with low buoyancy into a fresher region with higher buoyancy. In this model, we assume that the system reaches a steady state for the flow, the buoyancy, and the temperature. We have neglected the feedback from porosity variation on the flow ($\Pi$ is a constant and is equal to $\Pi_0$) which allows for such a quasi steady state with slowly evolving porosity and non-zero $\partial_t \phi$. Hence the neglect of transient porosity changes should be reasonable at early times, but will eventually break down at longer time as the accumulated porosity change becomes large. Due to the different behaviours in the aforementioned regions, there are presumably two different time scales according to which the porosity evolves. In region II, accounting for the narrow vertical boundary layer thickness, we get a melting rate $\partial_t \hat{\phi} \sim - Ra^{1/3} \Omega^{1/3} / \mathscr{C}$. In the corner region IV, using that $b\sim \partial_z p \sim Ra^{3/4}$, and $\nabla^2 b \sim b/h^2$, we obtain a melting rate $\partial_t \hat{\phi} \sim - Ra^{3/4} \Omega^{1/2} / \mathscr{C}$. This suggests a larger melting rate in the top region, in agreement with our observations.
	
\begin{figure}
			\centering
			\includegraphics[scale=1]{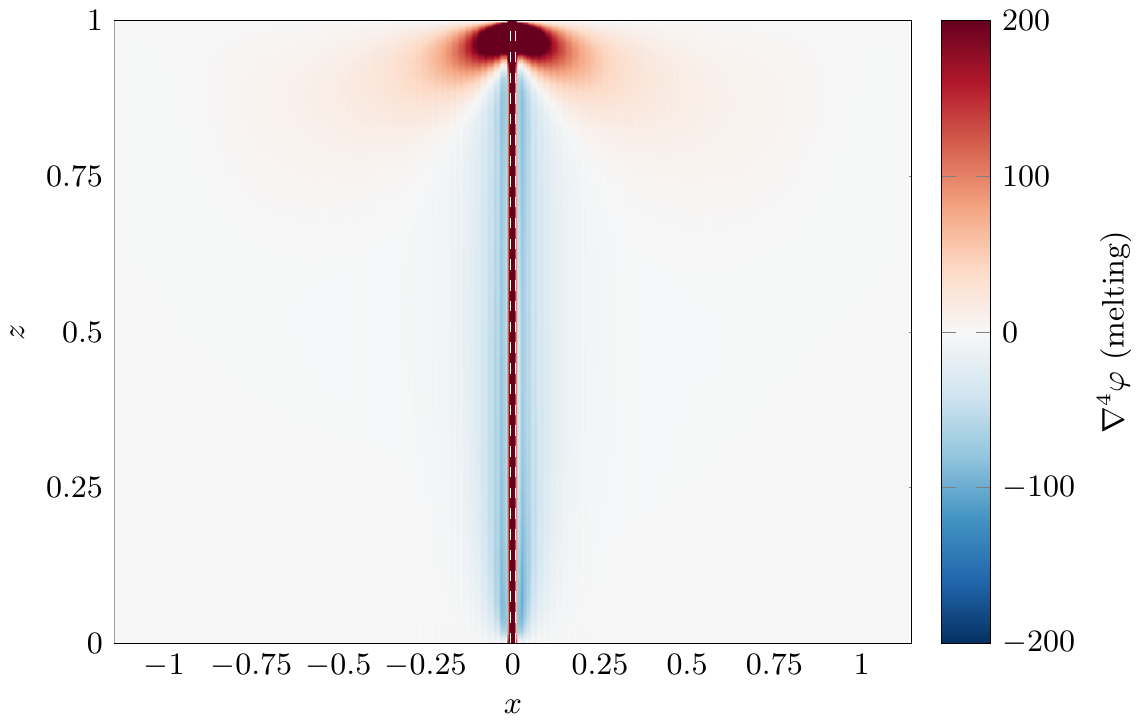}
			\caption{Dissolution rate, or solid fraction change, from the numerical simulation (red shows melting). The white dotted lines show the locations $\pm \sigma$ used for the Gaussian heat flux, at $Ra=200$ and $\Omega=1$.}
			\label{fig:ccl2}
		\end{figure}

		The instantaneous melting rate predicted by equation~\eqref{eq:melting} using the DNS is presented in figure~\ref{fig:ccl2}, in which areas with decreasing solid fraction are shown in red. This indicates instantaneous melt rate deviation away from steady state. Negative value for $\partial_t \hat{\phi}$ means that the solid fraction is decreasing, corresponding to ice melting. A strong and localised region of solid dissolution is identified at the top of the crack, corresponding to region IV, with a vertical extent of order $h$, which confirms the importance of the top region in the melting problem. Such dissolution is cause by upward advection of solute and a net convergence of the solute flux in the diffusive boundary layer in the upper corner region. An additional region of freezing can be seen close to the wall, next to the heated melting region which corresponds to the tendency for advection by the rising plume to deplete the initial background of solute and cause freezing. Because the solution predicts flow through regions of instantaneous freezing, it would only be valid for a system that initially has non-zero porosity and cannot be applied directly to an initial condition with pure eutectic solid. This freezing is a slower process than the melting, as $\partial_t \hat{\phi}$ varies over more than an order of magnitude between the freezing and the melting region at the crack, and even more if we look at the top region. On longer timescales, it is likely that this behaviour would be suppressed with a feedback of the porosity on the flow, as a solid fraction equal to $1$ would prevent the liquid from flowing. The relevent scalings, however, are still a good approximation of what is happening for a system starting with non-zero porosity at early times. Understanding the impact on porosity-dependent feedbacks on permeability, and evolution from an initially solid state represent interesting challenges for future work.

% ################################################################################ %
\section{Conclusions and Discussion}
\label{sec:conclusion}

	In this study, we developed a simple model of a buoyant plume generated by a heated wall with constant heat flux in an ideal mushy zone using a near-eutectic limit. We identified four different regions created along the wall through the heating process, and we show that the relevent part that dominates the melting is located near the surface over a short depth $h \propto Ra^{-1/2}$ (to within a multiplicative factor depending on the modified heat capacity $\Omega$). The proposed mechanism is robust, with local scalings holding for a wide range of Rayleigh number, and the different scalings observed in Direct Numerical Simulations agree well with the theory that uses asymptotics and self-similar solutions. From top to bottom these demonstrate an isotropic diffusive region I with linear scalings for buoyancy and velocity with height develops into a buoyant self-similar rising plume region II which scalings are given by a vertical boundary layer theory. Region II is connected to a top region IV through an isotropic region III. Region IV has a vertical velocity scaling consistent with a stagnation point flow, and the pressure gradient scaling imposed by an adjacent horizontal boundary layer that develops sidewards with approximate hydrostatic balance in the vertical.
	
	Some empirically diagnosed scalings still remain to be explained, however. In region III there are two different scalings for the buoyancy and the vertical velocity that are not theoretically explained. In this region, it is very likely that the vertical velocity matches with region II whereas the buoyancy and the pressure fields match with region IV, and that the total heat flux is conserved, but we have no supporting mathematical theory. No analytical result for region I, at the bottom of the crack, has been properly derived. The top region IV could be better understood by defining regions IV and V: one being the top corner, the other one being the horizontal boundary layer that extends near the surface in the $x$ direction. Different scalings can then be tested in region V accordingly to the horizontal boundary layer theory from section~\ref{sec:HBL}.
	
	The vertical extension of the top region can be estimated through the Rayleigh number, as $Ra= (H / h) ^{2}$. From the previous derivation, noting that the relevant temperature scale in this problem is $\Delta T = FH/C_p \kappa$, the Rayleigh number can be expressed into a porous media Rayleigh number as
		\begin{equation}
			Ra = \frac{g\alpha \Delta T \Pi_0 H}{\kappa \nu}
		\end{equation}
		In the case of icy satellites (e.g. Encaladus or Europa), we treat the planetary ice shell as having sea ice material properties and lower gravity on account of planetary size. Hence, due to the range of values for the different parameters involved in the problem if we estimate orders of magnitude for Enceladus such as $g=10^{-1}\mathrm{~m\cdot s^{-2}}$, $\alpha = 10^{-2}\mathrm{~K^{-1}}$, $\kappa = 10^{-7} \mathrm{~m^2 \cdot s^{-1}}$, $\nu = 10^{-6} \mathrm{~m^2 \cdot s^{-1}}$, $\Delta T \sim 10^2 \mathrm{~K}$, $\Pi_0 \sim 10^{-14} \mathrm{~m^2}$ (lower value than used for terrestrial sea ice (e.g. $\Pi_0 = 10^{-10} -10^{-12} \mathrm{~m^2}$ in \cite{polashenski2017}) since it is a colder and less open ice), and $H \sim 10^3-10^4\mathrm{~m}$, we obtain a relatively small Rayleigh number $Ra$, about $10$-$100$. A simple estimate indicates that the top region, in which large vertical pressure gradients are predicted, has a small extension of order $\mathcal{O}(h) \ll H$ and that the high melting region is therefore located near the surface.
	
	Increasing the Rayleigh number further triggers an instability near the wall (see figure~\ref{fig:psi}) and the present model of steady flow is not valid anymore. Starting from the linear steady state, such an instability could occur with an intense and short lived forcing generating large heat fluxes, which might be enough to create strong intermittent melting spots close to the heated wall. This unstable regime is very sensitive to slight changes in Rayleigh number (see figure~\ref{fig:psi}) and its threshold remains to be determined, which will be the focus of future studies.
	
	The geophysical setting of heated cracks in a planetary ice shell contains several complexities not included in the present model. A more accurate description of the problem could be derived, relaxing the constant permeability assumption. The flow would then feel a feedback from the melting of the mush and the phase change itself, as the permeability would be a function of the solid fraction. The geometry of the mushy region generated by melting may also introduce added complexity, as there may be an interface between the mushy region and the pure ice region, that can evolve through time as the mush is growing rather than the rectangular cavity considered here. Provided the mushy region extends beyond the extent of the boundary layers identified here, the present analysis should provides insights relevant to the more complex setting for the far-field flow.

% ################################################################################ %

\bigskip

\textbf{Acknowledgments}
	\\ The authors are grateful for the NSF OCE-1829864 support during the 2019 GFD Summer School at WHOI. SB also thanks IDEX Lyon for additional travel funds.
	
% ################################################################################ %

\bigskip
\textbf{Declaration of interests}
	\\ The authors report no conflict of interest.
	
% ################################################################################ %

%
\bigskip 
\bibliographystyle{jfm}
\bibliography{biblio-mushygeysers}

\begin{thebibliography}{24}
\expandafter\ifx\csname natexlab\endcsname\relax\def\natexlab#1{#1}\fi
\def\au#1{#1} \def\ed#1{#1} \def\yr#1{#1}\def\at#1{#1}\def\jt#1{\textit{#1}}
  \def\bt#1{#1}\def\bvol#1{\textbf{#1}} \def\vol#1{#1} \def\pg#1{#1}
  \def\publ#1{#1}\def\arxiv#1{#1}\def\org#1{#1}\def\st#1{\textit{#1}}

\bibitem[Anderson \& Guba(2019)]{anderson2019}
{\sc \au{Anderson, D.M.} \& \au{Guba, P.}} \yr{2019}  \at{Convective phenomena
  in mushy layers}.  \jt{Annual Review of Fluid Mechanics}  \bvol{52},  \pg{93
  -- 119}.

\bibitem[Bloomfield \& Huppert(2003)]{bloomfield2003}
{\sc \au{Bloomfield, L.J.} \& \au{Huppert, H.E.}} \yr{2003}  \at{Solidification
  and convection of a ternary solution cooled from the side}.  \jt{Journal of
  Fluid Mechanics}  \bvol{489},  \pg{269 -- 299}.

\bibitem[Burns {\em et~al.\/}(2020)Burns, Vasil, Oishi, Lecoanet \&
  Brown]{burns2019}
{\sc \au{Burns, K.~J.}, \au{Vasil, G.~M.}, \au{Oishi, J.~S.}, \au{Lecoanet, D.}
  \& \au{Brown, B.~P.}} \yr{2020}  \at{Dedalus: A flexible framework for
  numerical simulations with spectral methods}.  \jt{Physical Review Research}
  \bvol{2},  \pg{023068}.

\bibitem[Carey \& Gebhart(1982)]{carey1982}
{\sc \au{Carey, V.P.} \& \au{Gebhart, B.}} \yr{1982}  \at{Transport near a
  vertical ice surface melting in saline water: some numerical calculations}.
  \jt{Journal of Fluid Mechanics}  \bvol{117},  \pg{379 -- 402}.

\bibitem[Cheng \& Minkowycz(1977)]{cheng1977}
{\sc \au{Cheng, P.} \& \au{Minkowycz, W.~J.}} \yr{1977}  \at{Free convection
  about a vertical flat plate embedded in a porous medium with application to
  heat transfer from a dike}.  \jt{Journal of Geophysical Research}  \bvol{82},
   \pg{7B0014}.

\bibitem[Copley {\em et~al.\/}(1970)Copley, Giamei, Johnoson \&
  Hornbecker]{copley1970}
{\sc \au{Copley, S.M.}, \au{Giamei, A.F.}, \au{Johnoson, S.M.} \&
  \au{Hornbecker, M.F.}} \yr{1970}  \at{The origin of freckles in
  unidirectionnally solidified castings}.  \jt{Metallurgical Transactions}
  \bvol{1},  \pg{2193 -- 204}.

\bibitem[Fowler(1985)]{fowler1985}
{\sc \au{Fowler, A.~C.}} \yr{1985}  \at{The formation of freckles in binary
  alloys}.  \jt{Journal of Applied Mathematics}  \bvol{35},  \pg{159 -- 174}.

\bibitem[Furumoto(1975)]{furumoto1975}
{\sc \au{Furumoto, A.S.}} \yr{1975}  \at{A systematic program for geothermal
  exploration on the island of {H}awaii}.  \jt{Annual International Meeting,
  Society of Exploration in Geophysics} .

\bibitem[Gaidos \& Nimmo(2000)]{gaidos2000}
{\sc \au{Gaidos, E.J.} \& \au{Nimmo, F.}} \yr{2000}  \at{Tectonics and water on
  {E}uropa}.  \jt{Nature}  \bvol{405},  \pg{637}.

\bibitem[Guba \& Worster(2006)]{guba2006}
{\sc \au{Guba, P.} \& \au{Worster, M.~G.}} \yr{2006}  \at{Free convection in
  laterally solidifying mushy regions}.  \jt{Journal of Fluid Mechanics}
  \bvol{558},  \pg{69 -- 78}.

\bibitem[Hammond(2019)]{hammond2019}
{\sc \au{Hammond, N.P.}} \yr{2019}  \at{Near-surface melt on {E}uropa: Modeling
  the formation and migration of brines in a dynamic ice shell}.  \jt{Lunar and
  Planetary Science Conference}  \bvol{2168},  \pg{6024}.

\bibitem[Han \& Showman(2008)]{han2008}
{\sc \au{Han, L.} \& \au{Showman, A.P.}} \yr{2008}  \at{Implications of shear
  heating and fracture zones for ridge formation on {E}uropa}.  \jt{Geophysical
  Research Letters}  \bvol{35},  \pg{L03202}.

\bibitem[Hunke {\em et~al.\/}(2011)Hunke, Notz, Turner \&
  Vancoppenolle]{hunke2011}
{\sc \au{Hunke, E.C.}, \au{Notz, D.}, \au{Turner, A.K.} \& \au{Vancoppenolle,
  M.}} \yr{2011}  \at{The multiphase physics of sea ice: a review for model
  developers}.  \jt{Cryosphere}  \bvol{5},  \pg{989 -- 1009}.

\bibitem[Huppert(1990)]{huppert1990}
{\sc \au{Huppert, H.E}} \yr{1990}  \at{The fluid mechanics of solidification}.
  \jt{Journal of Fluid Mechanics}  \bvol{212},  \pg{209 -- 240}.

\bibitem[Huppert \& Worster(2012)]{huppert2012}
{\sc \au{Huppert, H.E.} \& \au{Worster, M.G.}} \yr{2012}  \at{Flows involving
  phase change}.  \bt{In {\em Environmental Fluid Dynamics Handbook\/} (ed.
  \ed{H.J. Fernando})}, chap.~35,  \pg{pp. 467 -- 477}.  \publ{CRC Press}.

\bibitem[Ingham \& Brown(1986)]{ingham1986}
{\sc \au{Ingham, D.~B.} \& \au{Brown, S.~N.}} \yr{1986}  \at{Flow past a
  suddenly heated vertical plate in a porous medium}.  \jt{Proceedings of the
  Royal Society of London. Series A}  \bvol{403},  \pg{51 -- 80}.

\bibitem[McCord {\em et~al.\/}(1999)McCord, Hansen, Matson, Johnoson, Crowlez,
  Fanale, Carlson, Smythe, Martin, Hibbitts, Granahan \& Ocampo]{mccord1999}
{\sc \au{McCord, T.B.}, \au{Hansen, G.B.}, \au{Matson, D.L.}, \au{Johnoson,
  T.V.}, \au{Crowlez, J.K.}, \au{Fanale, F.P.}, \au{Carlson, R.W.}, \au{Smythe,
  W.D.}, \au{Martin, P.D.}, \au{Hibbitts, C.A.}, \au{Granahan, J.C.} \&
  \au{Ocampo, A.}} \yr{1999}  \at{Hydrated salt minerals on {E}uropa's surface
  from the {G}alileo near-infrared mapping spectrometer ({NIMS})
  investigation}.  \jt{Journal of Geophysical Research}  \bvol{104},  \pg{11827
  -- 11851}.

\bibitem[Nimmo {\em et~al.\/}(2007)Nimmo, Spencer, Pappalardo \&
  Mullen]{nimmo2007}
{\sc \au{Nimmo, F.}, \au{Spencer, J.~R.}, \au{Pappalardo, R.~T.} \& \au{Mullen,
  M.~E.}} \yr{2007}  \at{Shear heating as the origin of the plumes and heat
  flux on {E}nceladus}.  \jt{Nature}  \bvol{447},  \pg{289 -- 291}.

\bibitem[Polashenski {\em et~al.\/}(2017)Polashenski, Golden, Perovich,
  Skyllingstad, Arnsten, Stwertka \& Wright]{polashenski2017}
{\sc \au{Polashenski, C.}, \au{Golden, K.~M.}, \au{Perovich, D.~K.},
  \au{Skyllingstad, E.}, \au{Arnsten, A.}, \au{Stwertka, C.} \& \au{Wright,
  N.}} \yr{2017}  \at{Percolation blockage: A process that enables melt pond
  formation on first year {A}rctic sea ice}.  \jt{Journal of Geophysical
  Research: Oceans}  \bvol{122},  \pg{1 -- 28}.

\bibitem[Tait \& Jaupart(1992)]{tait1992}
{\sc \au{Tait, S.} \& \au{Jaupart, C.}} \yr{1992}  \at{Compositional convection
  in a reactive crystalline mush and melt differentiation}.  \jt{Journal of
  Geophysical Research}  \bvol{97},  \pg{6735 -- 6756}.

\bibitem[Wells {\em et~al.\/}(2019)Wells, Hitchen \& Parkinson]{wells2019}
{\sc \au{Wells, A.~J.}, \au{Hitchen, J.~R.} \& \au{Parkinson, J.~R.~G.}}
  \yr{2019}  \at{Mushy-layer growth and convection, with application to sea
  ice}.  \jt{Philosophical Transactions A}  \bvol{377},  \pg{20180165}.

\bibitem[Worster(1986)]{worster1986}
{\sc \au{Worster, M.~G.}} \yr{1986}  \at{Solidification of an alloy from a
  cooled boundary}.  \jt{Journal of Fluid Mechanics}  \bvol{167},  \pg{481 --
  501}.

\bibitem[Worster(1997)]{worster1997}
{\sc \au{Worster, M.~G.}} \yr{1997}  \at{Convection in mushy layers}.
  \jt{Annual Reviews of Fluid Mechanics}  \bvol{29},  \pg{91 -- 122}.

\bibitem[Worster(2000)]{worster2000}
{\sc \au{Worster, M.~G.}} \yr{2000}  \at{Solidification of fluids}.  \bt{In
  {\em Perspectives in Fluid Dynamics\/} (ed. \ed{G.~K. Batchelor, H.~K.
  Moffatt \& M.~G. Worster})}, chap.~8,  \pg{pp. 393--444}.  \publ{Cambridge
  University Press}.

\end{thebibliography}

\end{document}